\documentclass[runningheads]{llncs}
\pdfoutput=1

\usepackage[T1]{fontenc}

\makeatletter
\RequirePackage[bookmarks,unicode,colorlinks=true,breaklinks=true]{hyperref}%
   \def\@citecolor{blue}%
   \def\@urlcolor{blue}%
   \def\@linkcolor{blue}%

\def\orcidID#1{\smash{\href{http://orcid.org/#1}{\protect\raisebox{-1.25pt}{\protect\includegraphics{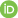}}}}}
\makeatother

\usepackage[noadjust]{cite}
\usepackage[dvipsnames,table]{xcolor}
\usepackage{graphicx}
\usepackage{xspace}

\usepackage{array}
\usepackage[inline]{enumitem}
\usepackage{amssymb}
\usepackage{amsmath}
\usepackage{cleveref}
\usepackage[normalem]{ulem}
\usepackage{bbding}
\usepackage{wrapfig}
\usepackage{multirow}
\usepackage{relsize}
\usepackage{booktabs}
\usepackage{ragged2e}
\usepackage{xifthen}
\usepackage{wasysym}
\usepackage{makecell}
\usepackage{slantsc}

\usepackage{tikz}
\usepackage{pgfplots}
\pgfplotsset{compat=1.8}
\usetikzlibrary{arrows,positioning}

\Crefname{figure}{Fig.}{Figs.}
\crefname{figure}{fig.}{figs.}
\Crefname{tabular}{Tab.}{Tabs.}
\crefname{tabular}{tab.}{tabs.}
\Crefname{section}{Sect.}{Sects.}
\crefname{section}{sect.}{sects.}
\Crefname{equation}{Eq.}{Eqs.}
\crefname{equation}{eq.}{eqs.}
\creflabelformat{equation}{#2#1#3}

\setitemize{noitemsep,topsep=0pt,parsep=0pt,partopsep=0pt,leftmargin=12.0pt}
\setenumerate{noitemsep,topsep=0pt,parsep=0pt,partopsep=0pt,leftmargin=12.5pt}
\setdescription{noitemsep,topsep=0pt,parsep=0pt,partopsep=0pt,leftmargin=12.5pt}

\newcolumntype{L}[1]{>{\RaggedRight\let\newline\\\arraybackslash}p{#1}}
\newcolumntype{C}[1]{>{\Centering\let\newline\\\arraybackslash}p{#1}}
\newcolumntype{R}[1]{>{\RaggedLeft\let\newline\\\arraybackslash}p{#1}}

\newcommand{\eg}{e.g.\ }
\newcommand{\ie}{i.e.\ }
\newcommand{\etal}{et al.\xspace}

\newcommand{\defeq}{\mathrel{\vbox{\offinterlineskip\ialign{\hfil##\hfil\cr{\tiny \rm def}\cr\noalign{\kern0.30ex}$=$\cr}}}}

\newcommand{\cm}{\checkmark}

\newcommand{\tool}[1]{\textsc{#1}\xspace}
\newcommand{\lang}[1]{\textsc{#1}\xspace}
\newcommand*{\jani}{\lang{Jani}}
\newcommand*{\modest}{\lang{Modest}}
\newcommand*{\dftres}{\tool{Dftres}}
\newcommand*{\epmc}{\tool{Epmc}}
\newcommand*{\fig}{\tool{Fig}}
\newcommand*{\gavs}{\tool{GAVS+}}
\newcommand*{\gist}{\tool{Gist}}
\newcommand*{\infamy}{\tool{Infamy}}
\newcommand*{\mcsta}{\tool{mcsta}}
\newcommand*{\modes}{\tool{modes}}
\newcommand*{\momba}{\tool{Momba}}
\newcommand*{\multigain}{\tool{MultiGain}}
\newcommand*{\multigaintwo}{\tool{MultiGain 2.0}}
\newcommand*{\toolset}{\tool{Modest Toolset}}
\newcommand*{\param}{\tool{Param}}
\newcommand*{\paynt}{\tool{Paynt}}
\newcommand*{\pet}{\tool{Pet}}
\newcommand*{\prism}{\tool{Prism}}
\newcommand*{\prismgames}{\tool{Prism-games}}
\newcommand*{\ragtimer}{\tool{Ragtimer}}
\newcommand*{\sequaia}{\tool{SeQuaiA}}
\newcommand*{\stamina}{\tool{Stamina}}
\newcommand*{\storm}{\tool{Storm}}
\newcommand*{\stormdft}{\tool{Storm-dft}}
\newcommand*{\stormdftres}{\tool{StormDftRes}}
\newcommand*{\tempest}{\tool{Tempest}}

\makeatletter%
\g@addto@macro\normalsize{%
  \setlength\abovedisplayskip{3pt}%
  \setlength\belowdisplayskip{3pt}%
  \setlength\abovedisplayshortskip{-3pt}%
  \setlength\belowdisplayshortskip{3pt}%
}%
\makeatother

\newcommand{\uu}{\underline{\hspace{0.4em}}}
\newcommand{\sgbest}[1]{{\bf{#1}}}
\newcommand{\sgtimeout}{{\emph{T/O}}}
\newcommand{\sgmemout}{{\emph{M/O}}}

\newcommand{\lrabest}[1]{{\bf{#1}}}

\definecolor{plotred}{RGB}{255,0,0}
\definecolor{plotgreen}{RGB}{0,255,0}
\definecolor{plotblue}{RGB}{0,0,255}
\definecolor{plotyellow}{RGB}{230,230,0}
\definecolor{plotcyan}{RGB}{0,255,255}
\definecolor{plotorange}{RGB}{255,127,0}
\definecolor{plotpink}{RGB}{255,0,255}
\definecolor{plotlightgray}{RGB}{192,192,192}
\definecolor{plotdarkgray}{RGB}{128,128,128}
\definecolor{plotdarkred}{RGB}{128,0,0}
\definecolor{plotgreenyellow}{RGB}{128,128,0}
\definecolor{plotdarkgreen}{RGB}{0,128,0}
\definecolor{plotpurple}{RGB}{128,0,128}
\definecolor{plotteal}{RGB}{0,128,128}
\definecolor{plotdarkblue}{RGB}{0,0,128}
\definecolor{plotlightred}{RGB}{205,92,92}
\definecolor{plotlightblue}{RGB}{176,196,222}

\newcommand{\quantileplotxlabel}{}
\newcommand{\quantileplotylabel}{}
\newlength{\quantileplotwidth}
\newlength{\quantileplotheight}
\newcommand{\quantileplotlegendcols}{1}
\newcommand{\quantileplot}[8]{%
	\begin{tikzpicture}	
	\begin{axis}[
	width=\quantileplotwidth,
	height=\quantileplotheight,
	unbounded coords=discard,filter discard warning=false, 
	xmin=#4,
	xmax=#5,
	ymin=#6,
	ymax=#7,
	ymajorgrids,
	ymode=log,
	axis x line=bottom,
	axis y line=left,
	ytick= {1, 10, 100, 1000,10000},
	yticklabels= {${\le}1$, $10$, $10^2$, $10^3$},
	xlabel=\quantileplotxlabel,
	ylabel=\quantileplotylabel,
	x label style={font=\scriptsize, yshift=5pt},
	y label style={font=\scriptsize, yshift=-3pt},
	yticklabel style={font=\scriptsize},
	scaled y ticks=false,
	xticklabel style={font=\scriptsize},
	legend columns=\quantileplotlegendcols,
	legend pos={#8},
	legend style={nodes={scale=0.75, transform shape},inner sep=1pt},
/pgfplots/legend image code/.code={\draw[mark repeat=2,mark phase=2,##1] plot coordinates {(0cm,0cm) (0.3cm,0cm)};}, 
	every axis plot/.append style={ultra thick},
	legend cell align={left}
	]
	\foreach \quantileplottool\quantileplotcolor in {#2}{%
		\edef\loopbody{
			\noexpand\addplot[\quantileplotcolor] table [x=i,y=\quantileplottool, col sep=tab] {#1};
		}
		\loopbody
	}
	\legend{#3}
	\end{axis}
	\end{tikzpicture}%
}

\definecolor{myred}{RGB}{158, 37, 16}
\definecolor{mygreen}{RGB}{8, 138,95}
\definecolor{mydarkblue}{RGB}{8,59,158}

\newcommand{\acronym}[1]{\ensuremath{\textsc{\larger{#1}}}\xspace}
\newcommand{\RE}{\acronym{re}}           
\newcommand{\RES}{\acronym{res}}         
\newcommand{\IS}{\acronym{is}}           
\newcommand{\ISPLIT}{\acronym{isplit}}   
\newcommand{\RESTART}{\acronym{restart}} 
\newcommand{\RESTARTPj}[1][j]{\ensuremath{\RESTART\acronym{\textsmaller{-}p}_{\!#1}}\xspace}
\newcommand{\PropP}[1]{\ensuremath{\mathtt{P_{=?}\boldsymbol[\,#1\,\boldsymbol]}}}
\newcommand{\PropS}[1]{\ensuremath{\mathtt{S_{=?}\boldsymbol[\,#1\,\boldsymbol]}}}
\newcommand{\PropF}[1][]{\ensuremath{F\ifthenelse{\equal{#1}{}}{}{_{#1}}~}}
\newcommand{\PropU}[1][]{\ensuremath{~U\ifthenelse{\equal{#1}{}}{}{_{#1}}~}}
\newcommand{\Prop}[1][]{\ensuremath{\varphi\ifthenelse{\equal{#1}{}}{}{_{#1}}}}
\newcommand{\Srop}[1][]{\ensuremath{\tilde{\varphi}\ifthenelse{\equal{#1}{}}{}{_{#1}}}}
\newcommand{\eminus}[1]{\ensuremath{\scalebox{.85}{\text{\scshape e-}#1}}\xspace}

\begin{document}

\title{%
Tools~at~the~Frontiers~of~Quantitative~Verification%
\thanks{%
The authors are ordered alphabetically.
This work was supported by
DFG RTG 2236/2 ({\scriptsize UnRAVeL})
and
DFG project TRR 248 ({\scriptsize CPEC}, ID 389792660),
by
the EU under MSCA grant agreements
101008233 ({\scriptsize MISSION}),
101034413 ({\scriptsize IST-BRIDGE}),
and
101067199 ({\scriptsize ProSVED}),
by
ERC Starting Grant 101077178 ({\scriptsize DEUCE}),
ERC Consolidator Grant 864075 ({\scriptsize CAESAR}),
and
ERC Advanced Grant 834115 ({\scriptsize FUN2MODEL}),
by
GA\v{C}R grant GA23-06963S ({\scriptsize VESCAA}),
by
National Science Foundation grant 1856733,
by
NextGenerationEU project D53D23008400006 ({\scriptsize SMARTITUDE}),
and
by NWO VENI grant 639.021.754.
}%
}
\subtitle{QComp 2023 Competition Report}
\author{
Roman~Andriushchenko\inst{1}\,\orcidID{0000-0002-1286-934X}
\and Alexander~Bork\inst{2}\,\orcidID{0000-0002-7026-228X}
\and Carlos~E.~Budde\inst{3}\,\orcidID{0000-0001-8807-1548}
\and Milan~\v{C}e\v{s}ka\inst{1}\,\orcidID{0000-0002-0300-9727}
\and Kush~Grover\inst{4}\,\orcidID{0000-0003-4575-1302}
\and Ernst~Moritz~Hahn\inst{5}\,\orcidID{0000-0002-9348-7684}
\and Arnd~Hartmanns\inst{5}$^{\text{\,\raisebox{-1pt}{\Envelope}}}$\,\orcidID{0000-0003-3268-8674}
\and Bryant~Israelsen\inst{6}\,\orcidID{0000-0002-9537-2645}
\and Nils~Jansen\inst{7}\,\orcidID{0000-0003-1318-8973}
\and Joshua~Jeppson\inst{6}\,\orcidID{0000-0002-8269-9489}
\and Sebastian~Junges\inst{7}\,\orcidID{0000-0003-0978-8466}
\and Maximilian~A.~K\"ohl\inst{8}\,\orcidID{0000-0003-2551-2814}
\and Bettina~K\"onighofer\inst{9}\,\orcidID{0000-0001-5183-5452}
\and Jan~K\v{r}et\'insk\'y\inst{4,10}\,\orcidID{0000-0002-8122-2881}
\and Tobias~Meggendorfer\inst{4,11,12}\,\orcidID{0000-0002-1712-2165}
\and David~Parker\inst{13}\,\orcidID{0000-0003-4137-8862}
\and Stefan~Pranger\inst{9}\,\orcidID{0009-0000-6011-9925}
\and Tim~Quatmann\inst{2}\,\orcidID{0000-0002-2843-5511}
\and Enno~Ruijters\,\orcidID{0000-0002-5855-5282}
\and Landon~Taylor\inst{6}\,\orcidID{0000-0002-4071-3625}
\and Matthias~Volk\inst{14}\,\orcidID{0000-0002-3810-4185}
\and Maximilian~Weininger\inst{4,11}\,\orcidID{0000-0002-0163-2152}
\and Zhen~Zhang\inst{6}\,\orcidID{0000-0002-8269-9489}
}
\authorrunning{R.\ Andriushchenko \etal} 

\institute{
Brno University of Technology, Brno, Czech Republic
\and RWTH Aachen University, Aachen, Germany
\and University of Trento, Trento, Italy
\and Technical University of Munich, Munich, Germany
\and University of Twente, Enschede, The Netherlands
\and Utah State University, Logan, Utah, USA
\and Radboud University, Nijmegen, The Netherlands
\and Saarland University, Saarland Informatics Campus, Saarbrücken, Germany
\and Graz University of Technology, Graz, Austria
\and Masaryk University, Brno, Czech Republic
\and Institute of Science and Technology Austria, Klosterneuburg, Austria
\and Lancaster University Leipzig, Leipzig, Germany
\and University of Oxford, Oxford, UK
\and Eindhoven University of Technology, Einhoven, The Netherlands
\\
\raisebox{-2pt}{\Envelope}~\email{a.hartmanns@utwente.nl}
}
\maketitle

\vspace{-17pt}
\begin{abstract}
The analysis of formal models that include quantitative aspects such as timing or probabilistic choices is performed by quantitative verification tools.
Broad and mature tool support is available for computing basic properties such as expected rewards on basic models such as Markov chains.
Previous editions of QComp, the comparison of tools for the analysis of quantitative formal models, focused on this setting.
Many application scenarios, however, require more advanced property types such as LTL and parameter synthesis queries as well as advanced models like stochastic games and partially observable MDPs.
For these, tool support is in its infancy today.
This paper presents the outcomes of QComp 2023:
a survey of the state of the art in quantitative verification tool support for advanced property types and models.
With tools ranging from first research prototypes to well-supported integrations into established toolsets, this report highlights today's active areas and tomorrow's challenges in tool-focused research for quantitative verification.
\end{abstract}

\section{Introduction}
\label{sec:Introduction}

The inclusion of quantitative aspects such as probabilistic choices, timing, and random delays in system modelling is crucial to ensure the correctness, performance, and dependability of the ever-increasing amount of complex safety- and economically-critical systems that support our societies.
Well-known examples include the use of randomised algorithms in Internet protocols to achieve both simplicity and scalability~\cite{KR01} or the fault tree modelling approach for safety assessment in the nuclear industry~\cite{GV81}.

Formally, these aspects can be captured in established mathematical \textbf{formalisms} like discrete- and continuous-time Markov chains (DTMCs and CTMCs) for probabilistic choices and stochastic timing, or more recent notions such as timed automata (TA)~\cite{AD94} for real-time behaviour.
Combining DTMCs with nondeterministic (\ie unquantified and controllable or adversarial) choices results in the nowadays-popular formalism of Markov decision processes (MDPs)~\cite{Bel57,Put94}.
These form the mathematical foundation of \emph{quantitative modelling}; for practical purposes, models are specified in a higher-level \textbf{modelling language}---such as \modest~\cite{BDHK06,HHHK13} or the \prism language~\cite{KNP11}---that is equipped with a semantics in terms of one of the formalisms.
When combined with a query for a numerical \emph{property} of a model, \eg for the probability of reaching a set of undesirable states or for the expected reward until a terminal state is reached, we have a basic \emph{quantitative verification} problem.

\subsection*{Basic Quantitative Verification Comparisons}

The \emph{basic problems}---\ie computing a
(i)~reachability probability,
(ii)~expected accumulated reward, or
(iii)~steady-state probability
on a DTMC, CTMC, or MDP model\footnote{Probabilistic timed automata (PTA)~\cite{KNSS02} can be turned into equivalent MDP~\cite{KNPS06,KNSW07} (or be solved as stochastic games~\cite{KNP09}), so treat them like MDP here.}---%
can be solved by various software tools developed over the past two decades.
Most tools use one of two approaches: either probabilistic model checking (PMC)~\cite{BAFK18,HJQW23}, which applies a numeric algorithm onto a complete in-memory representation of a model's state space, or statistical model checking (SMC)~\cite{AP18,LLTYSG19,YS02}, which randomly samples (or: \emph{simulates}) and statistically analyses a set of model behaviours; or a hybrid approach combining aspects of PMC and SMC such as partial exploration~\cite{KM20}, probabilistic planning~\cite{KSHH20,MK12}, (deep) reinforcement learning~\cite{BCCFKKPU14,GHJKS20}, or Monte Carlo tree search~\cite{ABKS18}.

\paragraph{The QComp competition.}
The 2019 Comparison of Tools for the Analysis of Quantitative Formal Models (QComp 2019)~\cite{HHHKKKPQRS19} compared the performance, versatility, and usability of nine such tools on a benchmark set of 100 basic quantitative verification problems\footnote{In the tool competition context, our verification problems are called \emph{benchmark instances}. Since most benchmark models are parametrised but basic problems ask for a single result value, a benchmark instance is a triple of a model, a concrete parameter valuation, and a property to evaluate. We cover parametric analysis in \Cref{sec:Parameters}.}.
It was the first tool competition in quantitative verification, part of the TOOLympics at TACAS 2019~\cite{BBBFGHHK19}.
The next edition of QComp in 2020~\cite{BHKKPQTZ20} used the same benchmark set, but focused more specifically on the different types of correctness guarantees provided by the different tools, highlighting the interplay between performance and precision in quantitative verification.
The results of QComp 2020 were presented at the ISoLA 2020/2021 conference.
Although the main outcomes of these two editions of QComp were performance results, they were meant as \emph{friendly competitions}:
We did not establish a ranking of tools or point out a ``winner''; rather, we highlighted the capabilities, strengths, and specific niches of all participating tools.
In particular, the results clearly showed that some tools were generalists solving many types of problems, while others were specialised to specific tasks where they performed much better than any other participant.
The entire performance evaluation and report-writing process was performed in close collaboration with the participants, most of which were the main developers of the respective tools.

\paragraph{Benchmark sets and formats.}
Aside from providing information about the capabilities and performance of the participating tools, these two editions of QComp also benefited the collaboration and alignment inside the quantitative verification research community:
In a parallel effort to QComp 2019, we established the Quantitative Verification Benchmark Set (QVBS)~\cite{HKPQR19}, from which the competition selected its 100 benchmark instances.
Although the QVBS' models were collected from various sources and came in various modelling languages, the QVBS as a matter of principle includes a translation of each model and its properties into the \jani interchange format~\cite{BDHHJT17}; as a result, any tool that supported \jani could participate in QComp 2019 and 2020.
\jani thus benefited QComp and the participating tool authors by simplifying frontend development, while QComp furthered the establishment of \jani as a community standard.

\subsection*{QComp 2023: Looking into the Future}

While improving solution methods for basic problems remains an active research topic (cf.\ \eg\cite{HH21,HK20,BJKKMS20,HJVMSB21,Har22,HJQW23}), most of today's work in quantitative verification focuses on what we refer to as \emph{advanced problems}:
Computing more complex properties on the basic models, computing basic properties on more complex models, or combinations thereof.
Most papers include an experimental evaluation, which, however, often uses an ad-hoc research prototype implementation, most of which are \emph{not} further developed into a stable and maintained tool.
Nevertheless, as QComp 2020 was presented at ISoLA 2020/2021, it became clear that more and more solution methods for advanced problems were being turned into tools of their own or integrated into existing stable tools such as \prism~\cite{KNP11} or \storm~\cite{HJKQV22}.
Therefore, the next edition of QComp that we present in this report, QComp 2023, shifts its focus towards these \emph{frontiers in quantitative verification}.

\paragraph{Aims.}
The aims of QComp 2023 are
(i)~to describe advanced problems in quantitative verification for which analysis algorithms have more recently been developed and first tool support is appearing,
(ii)~to document the state of the art of this tool support, in terms of what is available today and what pieces are still missing, and
(iii)~to perform the first comparative tests of these tools where appropriate.
The outcome of QComp 2023 is this competition report, which can serve as a guide to state-of-the-art tools for the domain expert faced with an advanced quantitative verification problem, as a historical reference for tool developers, and as a call to action pointing researchers to where better algorithms are still needed and tool developers to where ``market opportunities'' exist that can be filled with new tools.

\paragraph{Setup and process.}
QComp 2023 is a more friendly competition than ever:
It started with an open call for participation to the quantitative verification community in summer 2022.
The interested participants then followed an iterative process of determining \emph{categories} (\ie advanced problem scenarios) of interest, which included identifying and contacting additional participants.
Out of the group of all participants, we then established category coordinators who would lead the process needed to achieve the aims of the competition in their category.
As the QComp categories covered all kinds of research and tooling maturity levels, part of the task of the category coordinators was to establish the scope and refine the concrete aims of their category.
In categories where several sufficiently stable tools already exist, coordinators could choose to include a performance evaluation, while more cutting-edge categories would focus on a description of the category, available approaches, and prior experimental results if available.
The category coordinators delivered the outcomes of their category to the overall QComp 2023 coordinator before summer 2023;
over that summer, we integrated all contributions into this report.

In this distributed and flexible approach where the competition is divided into sub-groups that establish the actual aims of their own, QComp 2023 was modelled after the ARCH-COMP friendly competition on verifying continuous and hybrid systems (see \href{https://cps-vo.org/group/ARCH/FriendlyCompetition}{cps-vo.org/group/ARCH/FriendlyCompetition}), which has been running on this model successfully for seven editions as of today since 2017~\cite{ARCH17}, with its latest edition concluded just before QComp 2023 this summer.

\section{Categories and Participants}

As a friendly competition, QComp 2023 was open to all interested parties for suggesting, coordinating, and participating in categories related to quantitative verification.
All participants of QComp 2023 are co-authors of this competition report.
The competition as a whole was coordinated by A.~Hartmanns.
Before presenting the results of the individual categories in the remainder of this report, we give an overview of QComp 2023's ten categories with credits to the respective organisers and participants, and present the participating tools.

\subsection{Categories}

\begin{description}[itemsep=3pt,topsep=3pt,labelsep=3.5pt]

\item[Infinite-state and population models]
(\textit{$\infty$-state}, \Cref{sec:InfiniteState}):
coordinated by Z.\ Zhang;
participants: M.\ \v{C}e\v{s}ka, E.\ M.\ Hahn, J.\ Jeppson.

\item[Long-run average rewards]
(\textit{LRA}, \Cref{sec:LRA}):
coordinated by K.\ Grover, J.\ K\v re\-t\'insk\'y, and M.\ Weininger;
participants: A.\ Hartmanns, T.\ Meggendorfer, and T.\ Quatmann.

\item[Linear temporal logic]
(\textit{LTL}, \Cref{sec:LTL}):
coordinated by J.\ K\v ret\'insk\'y and M.\ Weininger.

\item[Multi-objective analysis]
(\textit{multi-obj.}, \Cref{sec:MultiObjective}):
coordinated by T.\ Quatmann;
participants: K.\ Grover, D.\ Parker, and M.\ Weininger.

\item[Parametric Markov models]
(\textit{parametric}, \Cref{sec:Parameters}):
coordinated by S.~Junges.

\item[Partially-observable MDPs]
(\textit{POMDPs}, \Cref{sec:POMDPs}):
coordinated by A.\ Bork;
participants: R.\ Andriushchenko and D.\ Parker.

\item[Rare events]
(\textit{rare events}, \Cref{sec:RareEvents}):
coordinated by C.\ E.\ Budde;
participants: B.\ Israelsen, E.\ Ruijters, L.\ Taylor, M.\ Volk, and Z.\ Zhang.

\item[Robust decision-making under uncertainty]
(\textit{uncertainty}, \Cref{sec:Uncertainty}):
coordinated by N.\ Jansen;
participants: D.~Parker.

\item[State space exploration]
(\textit{exploration}, \Cref{sec:StateSpaceExploration}):
coordinated by M.\ A.\ K\"ohl;
participants: A.\ Hartmanns and T.\ Quatmann.

\item[Stochastic games]
(\textit{st.\ games}, \Cref{sec:StochasticGames}):
coordinated by D.\ Parker;
participants: B.\ K\"onighofer, T.\ Meggendorfer, S.\ Pranger, and M.\ Weininger.

\end{description}

\begin{table}[t]
\caption{Tools participating in QComp 2023's different categories}
\label{tab:ParticipatingTools}
\centering
\setlength{\tabcolsep}{6pt}
\begin{tabular}{lcccccccccl}
\toprule
& \rotatebox{90}{$\infty$-state} & \rotatebox{90}{LRA} & \rotatebox{90}{LTL} & \rotatebox{90}{multi-obj.} & \rotatebox{90}{parametric} & \rotatebox{90}{POMDPs} & \rotatebox{90}{rare events} & \rotatebox{90}{uncertainty} & \rotatebox{90}{exploration} & \rotatebox{90}{st.~games}\\
\midrule
\textit{experiments} 
             &   N   &   Y   &   N   &   Y   &   N   &   Y   &   Y   &   N   &   Y   &   Y   \\
\textit{benchmarks}
             &  --   &  20   &  --   &  66   &  --   &   3   &  10   &  --   &  229  &  16   \\
\midrule
\dftres      &       &       &       &       &       &       &  \cm  &       &       &       \\
\epmc        &       &       &  \cm  &  \cm  &  \cm  &       &       &       &       &  \cm  \\
\fig         &       &       &       &       &       &       &  \cm  &       &       &       \\
\infamy      &  \cm  &       &       &       &       &       &       &       &       &       \\
\mcsta       &       &  \cm  &       &       &       &       &       &       & \multirow{2}{*}{\cm} & \\
\modes       &       &       &       &       &       &       &  \cm  &       &       &       \\
\momba       &       &       &       &       &       &       &       &       &  \cm  &       \\
\multigain   &       &  \cm  &  \cm  &  \cm  &       &       &       &       &       &       \\
\param       &       &       &       &       &  \cm  &       &       &       &       &       \\
\paynt       &       &       &       &       &       &  \cm  &       &       &       &       \\
\pet         &       &  \cm  &       &       &       &       &       &       &       &  \cm  \\
\prism       &       &       &  \cm  &  \cm  &  \cm  &  \cm  &       &  \cm  &       &       \\
\prismgames  &       &  \cm  &       &  \cm  &       &       &       &       &       &  \cm  \\
\ragtimer    &       &       &       &       &       &       &  \cm  &       &       &       \\
\sequaia     &  \cm  &       &       &       &       &       &       &       &       &       \\
\stamina     &  \cm  &       &       &       &       &       &       &       &       &       \\
\storm       &       &  \cm  &  \cm  &  \cm  &  \cm  &  \cm  &       &       &  \cm  &       \\
\stormdftres &       &       &       &       &       &       &  \cm  &       &       &       \\
\tempest     &       &  \cm  &       &       &       &       &       &       &       &  \cm  \\
\bottomrule
\end{tabular}
\end{table}

\subsection{Participating Tools}

Various tools ranging from research prototypes to mature toolsets are available today to tackle the problems covered by the different categories.
In \Cref{tab:ParticipatingTools}, we list which tools participated in which of the categories of QComp 2023.
The meaning of ``participate'', however, can have a very different meaning in different categories; for example, the \textit{parametric} category only names the four tools that support the analysis of parametric Markov models, while the \textit{multi-obj.}\ category benchmarks its five participating tools and reports on their relative performance.
Categories that include an experimental evaluation such as runtime benchmarking are indicated by a ``Y'' in the row labelled ``\textit{experiments}''; then row ``\textit{benchmarks}'' states the number of benchmark instances considered in the experimental evaluation\footnote{More benchmarks may be \emph{available} for the problems covered by a category, and category \textit{parametric} has no performance evaluation but introduces a benchmark set.}.
Participation of a tool in any category was voluntary and not automatic; in particular, if a tool does not participate in a certain category, this does \emph{not} imply absence of support for the advanced properties or model types that the category focusses on in the tool.
To allow the individual category sections to focus on the specifics of their topic, we briefly introduce all 19 tools:

\begin{description}[itemsep=3pt,topsep=3pt,labelsep=-\fontdimen2\font]

\item[\dftres]~\cite{BRS20},
available at \href{https://github.com/utwente-fmt/DFTRES}{github.com/utwente-fmt/DFTRES},
is a statistical model checker designed for repairable dynamic fault trees (DFTs~\cite{RRBS19}) specified in Galileo and more general CTMCs specified in \jani.
It is written in Java and is portable to, at least, Linux, Windows, and macOS.

\item[\epmc]~\cite{FHLSSTZ22},
available at \href{https://github.com/iscas-tis/ePMC}{github.com/iscas-tis/ePMC},
is an extensible probabilistic model checking framework mostly written in Java.
It is a successor of \textsc{IscasMC}~\cite{HLSTZ14}.

\item[\fig]~\cite{BDMS22},
available at \href{https://git.cs.famaf.unc.edu.ar/dsg/fig}{git.cs.famaf.unc.edu.ar/dsg/fig},
is a statistical model check\-er for transient and steady state reachability properties in CTMCs and input/output stochastic automata (IOSA)~\cite{DM18}.
\fig is written in C++ and runs on Linux.

\item[\infamy]~\cite{HHWZ09},
available at \href{https://depend.cs.uni-saarland.de/tools/infamy/}{depend.cs.uni-saarland.de/tools/infamy},
is a tool with the purpose of model checking formulae in continuous stochastic logic (CSL)~\cite{ASSB00,BHHK03} on infinite-state CTMC specified in a variant of the \prism language by exploring the model up to a certain depth repeatedly.
\infamy can also handle certain reward properties.

\item[\mcsta],
available at \href{https://www.modestchecker.net/}{modestchecker.net},
is the explicit state model checker of the \toolset~\cite{HH14}, a collection of tools for the modelling and analysis of stochastic timed and hybrid systems.
Its core functionality is the disk-based explicit-state model checking of MDPs~\cite{HH15}, MAs~\cite{BHH21}, PTAs~\cite{HH09}, and stochastic timed automata~\cite{HHH14}.
The \toolset is mainly written in C\# and runs on 64-bit Linux, macOS, and Windows systems.
It supports the Modest~\cite{BDHK06,HHHK13} and \jani~\cite{BDHHJT17} input languages.

\item[\modes]~\cite{BDHS20},
available at \href{https://www.modestchecker.net/}{modestchecker.net},
is the \toolset's statistical model checker.
It supports the same input languages and platforms as \mcsta.
It contains simulation engines specialised to different formalisms from DTMCs to stochastic hybrid automata with general probability distributions (SHA)~\cite{FHHWZ11}, including support for non-linear continuous dynamics~\cite{NHR21}.

\item[\momba]~\cite{KKH21}, available at \href{https://momba.dev}{momba.dev}, 
is a Python library centred around \jani with the goal of providing easy access to quantitative modelling capabilities.

\item[\multigain]~\cite{BCFK15}
is an extension of \prism for multiple long-run average rewards.
\multigaintwo~\cite{BEGKW23}, available at \href{https://zenodo.org/records/10610642}{zenodo.org/records/10610642}, builds on \multigain, adding support for verification and strategy synthesis for~LTL.

\item[\param]~\cite{HHZ11},
available at \href{https://depend.cs.uni-saarland.de/tools/param/}{depend.cs.uni-saarland.de/tools/param},
was the first tool implementing verification algorithms for parametric Markov models.

\item[\paynt]~\cite{ACJKS21}, available at \href{https://github.com/randriu/synthesis}{github.com/randriu/synthesis},
is a tool originally developed for the inductive synthesis of probabilistic programs.
It aims at directly synthesising finite-state controllers for partially-observable MDPs.

\item[\pet]~\cite{Meg22}
available at \href{https://gitlab.lrz.de/i7/partial-exploration}{gitlab.lrz.de/i7/partial-exploration},
is a model checker focusing on value iteration approaches augmented by partial exploration, based on~\cite{BCCFKKPU14} for reachability with subsequent extensions to mean payoff~\cite{ACDKM17} and cores~\cite{KM20}.
It is backed by tailored data structures and algorithms for this purpose, and implemented in Java.

\item[\prism]~\cite{KNP11},
available at \href{https://www.prismmodelchecker.org/}{prismmodelchecker.org},
is a widely-used probabilistic model checker supporting a large range of models and temporal logics.
It is a user-friendly tool that comes with a cross-platform graphical user interface.
\prism is mostly written in Java, with some algorithms implemented in C.

\item[\prismgames]~\cite{KNPS20},
available at \href{https://www.prismmodelchecker.org/games/}{prismmodelchecker.org/games},
is an extension of \prism focused on the verification of stochastic games.

\item[\ragtimer]~\cite{ITZ23,TIZ23},
available at \href{https://github.com/fluentverification/ragtimer}{github.com/fluentverification/ragtimer},
is designed for chemical reaction networks (CRNs) modeled as CTMCs, combining guided stochastic simulation and commutability properties to compute lower-bound rare event probabilities from a partial state space.

\item[\sequaia]~\cite{CCK20},
available at \href{https://sequaia.model.in.tum.de/}{sequaia.model.in.tum.de},
offers two powerful engines for the quantitative analysis of population models given as chemical reaction networks via abstraction and simulation.
Both build on an interval population abstraction of the underlying CTMC.
\sequaia comes with a GUI, allowing for convenient modelling and tweaking the models as well as displaying the abstractions and analyses results for better explainability.

\item[\stamina]~\cite{JVIRWBMZWZ23,RNBMZ22,NMMZZ19,NZMZM19},
available at \href{https://staminachecker.org}{staminachecker.org},
is an infinite-state PMC tool that iteratively explores a partial state space for a bounded or unbounded CTMC model.
The CTMC transient analysis on the partial state space is delegated to \prism's and \storm's PMC engines.
\stamina/\prism
implements the \stamina 2.0 algorithm and interfaces with \prism's Java API and uses \prism for model parsing and checking.
\stamina/\storm is a reimplementation and extension of the \stamina 2.0 algorithm using \storm. 

\item[\storm]~\cite{HJKQV22},
available at \href{https://www.stormchecker.org/}{stormchecker.org},
is a general purpose, high-per\-for\-mance feature-rich probabilistic model checker built around a modular core with an emphasis on time and memory efficiency.
Written in C\texttt{++}, \storm's modular design enables the utilization of different model checking engines catering to the characteristics of different models.
Notably, \storm excels in efficient symbolic model checking through its \texttt{dd} engine leveraging binary decision diagrams (BDDs).

\item[\stormdftres],
available at \href{https://gitlab.utwente.nl/fmt/fault-trees/storm-dft-res}{gitlab.utwente.nl/fmt/fault-trees/storm-dft-res},
implements multi-threaded Monte Carlo simulation for (non-repairable) DFTs given in either the Galileo or a custom format.
\stormdftres builds on the \stormdft library~\cite{VJK18} of \storm, which implements efficient state space generation for DFTs by exploiting \eg irrelevant failures and symmetries.

\item[\tempest]~\cite{PKPB21}
available at \href{https://tempest-synthesis.org/}{tempest-synthesis.org}, is based on the \storm model checker, extending its feature set to turn-based stochastic games with a focus on synthesizing most-permissive strategies.

\end{description}

\section{Infinite-State and Population Models}
\label{sec:InfiniteState}

In many biochemical reaction and synthetic biology applications, very complex systems are studied and thus software tools become very advantageous and even indispensable for their understanding.
For instance, the signalling pathways, chemical reaction networks, and genetic regulatory networks under study consist of many concurrent reactions running at very different speeds and probabilities, with species of both low and high copy numbers.
This results in stiff systems suffering from stochasticity/multi-modality and state-space explosion, respectively~\cite{Kam92,Gou05}, calling for dedicated analysis tools.

In order to analyse such systems, so-called \emph{population models} are considered.
A state of a population model is a tuple of integers, with the $i$-th component representing the copy number of the $i$-th species.
Hence the state space is typically (countably) infinite.
Transitions between states represent executing one reaction of the system.
Given that the timing aspect is crucial and that the probability for a reaction to occur is (approximately) exponentially distributed as a function of real time, the model can be defined as a CTMC.
This explicit model can be derived directly from a symbolic representations of the system as, say, a \emph{chemical reaction network} (CRN):
The rates of the CTMC can be computed from the rates of the CRN reactions and the copy numbers in each state using the mass action kinetics.\footnote{Consequently, more symbolic models such as stochastic Petri nets are hard to use since the rate of a transition for a particular reaction differs from state to state.}
This transformation immediately enables the applicability of probabilistic model checkers for CTMC to biological systems.
However, in order to make the analysis practical, the population structure has to be exploited in dedicated ways.
In particular, one has to deal with the huge and in general infinite state spaces.

To handle such state spaces, various \textbf{reduction techniques} have been proposed that either truncate states of the underlying CTMC with insignificant probability~\cite{MK06} or leverage structural properties of the CTMC to aggregate/\allowbreak{}lump selected sets of states~\cite{AACK21,BBGW21}.
The \emph{interval abstraction} of the species population is a widely used approach to mitigate the state-space explosion problem~\cite{ZWC09}.
Alternatively, several hybrid models have been considered, such as
treating only small-population species stochastically while using a deterministic semantics for large-population species~\cite{HMMW10},
applying a moment-based description for medium/high-population species~\cite{HWKT13}, or
using the LNA approximation with an adaptive partitioning of the species according to leap conditions~\cite{CKL16}.

The investigated \textbf{properties} range from transient (``What is the (distribution over) states at time $t$?'') to steady-state analysis (concerning the limiting distribution or LRA reward). 
The typical output of a tool for such a query is either a a certain probability bound or an exact probability (or probability bound) of the predicate being true.
Given the numeric character of the results and methods, approximate solutions are considered.
Further, in contrast to verification, given that the systems are mostly neither safety-critical, nor completely modelled, it is typically acceptable to produce results without precise error bounds:
often by simulation-based techniques~\cite{Gil77} or aggressively practical, \eg semi-quantitative~\cite{CK19}, model-based approaches.

\subsection{Tool Support and Benchmarks}

\begin{table}[t]
\caption{Feature comparison of tools for population models}
\label{tab:tools_Inf}
\centering
\setlength{\tabcolsep}{4pt}
\begin{tabular}{lccccc}
\toprule
Tool & Platforms
& Approach
& Models & Syntax
& Semantics
\\
\midrule
			\infamy & Linux
			& \makecell[c]{model checking\\+ state truncation}
				& CTMC & \prism
& CTMC
\\[8pt]
	\sequaia & \makecell[c]{multi-\\platform}
& \makecell[c]{population abstraction\\+ numerical, simulation}
& \makecell[c]{population\\models} & \makecell[c]{GUI,\\dedicated}
& CTMC
\\[8pt]
			\stamina & \makecell[c]{Linux,\\macOS}
& \makecell[c]{model checking\\+ state truncation}
& CTMC & \prism
& CTMC
\\
\bottomrule
\end{tabular}
\end{table}

The main technical characteristics of the available tools participating in this category of QComp 2023 are listed in \Cref{tab:tools_Inf}.
\begin{description}[itemsep=3pt,topsep=3pt,labelsep=0pt]

\item[\infamy]
model-checks infinite-state CTMC specified in a variant of the \prism language.
It is capable of handling the time-bounded subclass of the logic CSL and certain reward properties.
It explores the model up to a certain depth repeatedly while descending into the nested CSL formula.
\infamy provides different means for finding a stopping criterion for the state-space exploration.
This is because there is a trade-off between when to stop and the memory needed to store the finite truncation of the state space.

\item[\sequaia]
offers two engines, both building on a ``population'' abstraction of the underlying CTMC, abstracting concrete copy numbers to given intervals.
The first one~\cite{CCK20} computes an abstraction of the CTMC using \emph{acceleration}, abstracting not only states and single transitions, but taking into account sequences of transitions.
The resulting model is
(i)~small enough to \emph{explain} the overall dynamics, and
(ii)~despite the induced imprecision, allows for a \emph{semi-quantitative analysis}, computing not the exact probabilities of different behaviours, but their orders of magnitude, which is often sufficient in the biological applications.
The engine thus features unprecedented scalability, analysing standard complex benchmarks within a fraction of a second, while it is precise enough to conclude on the qualitative behaviour of the system including rare behaviours and on rough estimates of the quantities (population sizes, times).
The second engine provides a more precise quantitative analysis by uniquely \emph{combining} the population abstraction with advanced simulation techniques~\cite{HCKM22}.
It is based on a memoization technique that combines previously generated \emph{segments} of runs defined over abstract states to generate new simulations more efficiently while preserving the original system dynamics and its diversity.
It adapts online to identify the most important abstract states and thus utilizes the available memory efficiently.
In combination with a novel fully automatic and adaptive hybrid simulation scheme, this speeds up the generation of trajectories and correctly predicts the transient behaviour of complex stochastic systems.

\item[\stamina]
iteratively explores a partial state space where a majority of the probability mass resides.
It expands the state space \emph{on the fly} based on the estimated state reachability probability, and truncates a search path when the estimate drops below a user-specified threshold.
\stamina then performs time-bounded transient PMC analysis by interfacing with \prism or \storm.
In this way, it computes a lower and upper bound, $P_{\!\mathit{min}}$ and $P_{\!\mathit{max}}$, respectively, such that the actual probability of the CSL property under verification lies within $[P_{\!\mathit{min}}, P_{\!\mathit{max}}]$.
The tightness of the probability window, $w = P_{\!\mathit{max}} - P_{\!\mathit{min}}$, is specified by the user, albeit with higher run-time for a smaller $w$.
\stamina can efficiently produce an accurate probability bound for CTMCs with an extremely large or infinite state space.
It is not restricted to specific types of input models as long as they can be modelled as CTMCs using the \prism modelling language.
Examples include genetic regulatory networks~\cite{FDBDVM20,MZRWM14}, biochemical reaction systems~\cite{DSFZ13,KM08}, dynamic fault trees (DFTs)~\cite{VJK18}, and queuing network models~\cite{HMKS99,Jac57}.
\stamina has been designed to support multiple exploration methods, and can be tailored to the model or property under verification.
It has also been designed to be user-friendly and modular.
Additionally, a graphical user-interface (written in Qt5) is under active development and will enhance user-experience and ease of use of \stamina.
\end{description}
Benchmarks include the CTMC models in the QVBS, the PRISM Benchmark Suite~\cite{KNP12}, and the \infamy case studies\footnote{\url{https://depend.cs.uni-saarland.de/tools/infamy/casestudies/}}. 
In addition, the Stochastic Model Case Studies repository~\cite{Buecherl2023}\footnote{\url{https://github.com/fluentverification/CaseStudies_StochasticModelChecking}} hosts a large collection of case studies focusing on biochemical systems with infinite state spaces.
\stamina has been evaluated on selected CTMC benchmarks from these benchmark suites, \eg\cite{JVIRWBMZWZ23,RNBMZ22}.
\sequaia has been evaluated on models describing challenging CRNs from the literature~\cite{CCK20}.

\subsection{Outlook}

Aside from the scalability limitation the tools are trying to mitigate, there are specific challenges in the analysis of biochemical and synthetic biological systems.

First, it is a very strong assumption that the correct model is available.
Consequently, the analysis methods should be able to effectively handle various forms of \textbf{model uncertainty}, including unknown reaction rate parameters, unknown reactants or products, as well as unknown species bounds.
The uncertainty can be modelled by various formalisms such as parametric or interval CTMC (see \Cref{sec:Uncertainty}), for which the existing tools offer only very limited support.

Second, \textbf{concurrency} is fundamental to these systems, as their constituent chemical reactions are often simultaneously enabled.
All enabled concurrent reactions may occur in a state but with very different probabilities, and their noisy operating environment can introduce extremely infrequent but potentially detrimental faults (see \Cref{sec:RareEvents}).
Additionally, their regulatory nature and constituent reversible reactions can cause \textbf{cyclic behaviours} and often require long reaction execution sequences to reach a desirable state.

Finally, the verification tools should offer to the users (\ie biologists) not only the verification result, but also an artifact in the form of a critical sub-system or a critical set of paths allowing the users to \textbf{interpret and explain} the results.
While some rudimentary effort has been made, \eg \cite{CK19}, this field is wide open.

\section{Long-Run Average Rewards}
\label{sec:LRA}

Many frequently-studied classes of properties of probabilistic systems are based on \emph{rewards}.
A reward function assigns to every state (or action or state-action pair) a number modelling a cost (or a payoff) related to the single move.
These rewards are accumulated over infinite paths in various manners.
Popular ways are discounted, total, and average rewards~\cite{Put94}.
While the \emph{discounted reward} is heavily used in diverse applications ranging from economics to robotics, and is very easy to optimize, it essentially reflects a limited time horizon only.
The \emph{total reward} can reflect longer horizons better (\eg unbounded reachability), yet not really the infinite-run behaviour.
The \emph{average reward} (also known as long-run average reward, limit-average reward, steady-state reward, or mean payoff) captures much more adequately the performance over an unknown or variable horizon (see \eg \cite{Sch93}).
Consequently, it is used to model \eg performance properties, such as the average delay between requests and responses, the average rate of a particular event, etc.
Considering the infinite horizon makes both classic and learning algorithms less efficient.
The whole problem is thus more difficult, and also less studied in the context of AI or robotics.
In contrast, it is significantly studied in formal verification where performance and dependability are critical and hard guarantees are desirable.
Related to the average reward and reducible to it are constraints on the steady state of a system, which become more studied also in the context of AI; see \eg \cite{AABTA22,Kre21}.
The algorithms for long-run average (LRA) reward properties again span the whole spectrum of linear and dynamic programming, including value and strategy iteration, with the usual advantages and disadvantages.
A specific case is the traditional steady-state analysis on (fully stochastic) Markov chains.
There, solving a system of linear equations is sufficient, but for efficiency reasons often replaced by value iteration, too.

\subsection{Algorithms and Tool Support}

\begin{table}[t]
\centering
\caption{Feature comparison of tools for average-reward properties}
\label{tab:mp-tools}
\setlength{\tabcolsep}{6pt}
\renewcommand*{\arraystretch}{1.1}
\begin{tabular}{llll} \toprule
Tool    &   Objective   &   Model   &   Guarantees\\
\midrule
\mcsta & ELRA & CTMC, MA & $\varepsilon$\\
\multigain & ELRA, SS & MDP & E-FP \\
\pet & ELRA & DTMC, CTMC, MDP & $\varepsilon$\\
\prismgames & ELRA & TSG & none \\
\storm & ELRA, SS & DTMC, CTMC, MDP, MA & $\varepsilon$, E-RA\\
\tempest & ELRA & TSG & none\\
\bottomrule
\end{tabular}
\end{table}

\Cref{tab:mp-tools} gives an overview of tools supporting average-reward properties, differentiating
the exact kind of supported objective (SS: steady-state or ELRA: expected long-run average reward),
the supported models (where MA are Markov automata~\cite{EHZ10} and TSG are turn-based stochastic games, see \Cref{sec:StochasticGames}), and
the guarantees provided on the precision of the result (either none, $\varepsilon$-precise, E-FP: exact up to floating-point precision, or ERA: exact using rational arithmetic).
We complement this high-level overview with short tool descriptions:
\begin{description}[itemsep=3pt,topsep=3pt,labelsep=0pt]

\item[\mcsta]
supports~\cite{BHH21} model-checking LRA reward properties in MA (and thus also in CTMC as a special case) using either a reduction to linear programming~\cite{GTHRS14} or an $\varepsilon$-precise method based on value iteration~\cite{BWH17}.

\item[\multigain]
implements a linear programming-based approach~\cite{BBCFK14} for multi-objective steady-state and LRA reward objectives in MDP (see also \Cref{sec:MultiObjective}).

\item[\pet]
focuses on partial-exploration techniques, for which it includes an extension to mean-payoff objectives~\cite{ACDKM17}.

\item[\prismgames]
supports a wide range of zero-sum properties; for LRA rewards (and its multi-objective variant of \emph{ratio objectives}), the stochastic games are required to be controllable multichains, \ie the sets of states that can occur in any maximal end component must be almost surely reachable.

\item[\storm]
can answer a wide range of queries for many different models, offering both value iteration-based approaches providing $\varepsilon$-precise results as well as linear programming-based algorithms using exact rational arithmetic.
We highlight that \storm is the only tool able to handle negative rewards directly, while others require the rewards to be rescaled first (see Appendix A in the full version of~\cite{KMW23} for the standard transformation).

\item[\tempest]
implements mean-payoff analysis for TSGs on top of \storm, using value iteration with explicit state space representations.

\end{description}

\subsection{Performance Comparison}

\begin{table}[t]
\caption{Performance comparison results of tools for average-reward properties}
\label{tab:mp-perf}
\centering
\setlength{\tabcolsep}{4pt}
\renewcommand*{\arraystretch}{1.1}
\begin{tabular}{lrrrrr}\toprule
Model (Parameters) & \# states & Value & ~~~~\storm & MG 2.0 & \pet\\
\midrule
busyRing                                 & 1,912     & 1.0   & \lrabest{0.04} &1.66 & 2.42 \\
coin (\texttt{N}=2, \texttt{K}=5)        & 656       & 1.0   & \lrabest{0.01} & 0.57 & 3.36\\
consensus (\texttt{N}=4, \texttt{K}=10)  & 104,576   & 1.0   & \lrabest{4.83} & 189.95 & 4,505.29\\
csma (\texttt{N}=2, \texttt{K}=4)        & 4,958     & 1.0   & \lrabest{0.04} & 1.52 & 4.49\\
cs\_nfail                                & 184       & 0.33  & \lrabest{0.02} & 0.48 & 2.85\\
eajs (\texttt{energy\_capacity}=500)      & 93,228    & 3.64  & \lrabest{0.36} & 11.02 & 168.80\\
eajs (\texttt{energy\_capacity}=1000)     & 193,728   & 3.64  & \lrabest{0.71} & 36.61 & 345.28\\
firewire (\texttt{deadline}=20, \texttt{delay}=2)
                                          & 2,862     & 0.0   & \lrabest{0.04} & 1.15  & 6.22\\
ij (10)                                     & 1,023     & 1.0   & \lrabest{0.02} & 0.83 & 3.61\\
investor                                  & 6,688     & 0.95  & \lrabest{0.07} & 4.19 & 5.15\\
mer (3)                                   & 15,622    & 1.5   & \lrabest{0.81} & 12.77 & 16.17\\
mer (4)                                   & 119,305   & 1.5   & 41.82 & 2,385.99 & \lrabest{41.52}\\
pacman (\texttt{MAXSTEPS}=10)             & 6,852     & 0.78  & \lrabest{0.16} & 3.41 & 4.71\\
pacman (\texttt{MAXSTEPS}=15)             & 96,894    & 0.99  & \lrabest{2.95} & 14.31 & 9.77\\
pnueli-zuck                               & 2,701     & 1.0   & \lrabest{0.04} & 1.31 & 3.91\\
rabin (3)                                 & 15,622    & 0.86  & \lrabest{0.23} & 11.08 & 8.25\\
sensors (\texttt{K}=5)                    & 267       & 0.45  & \lrabest{0.01} & 0.60 & 1.97\\
virus (3)                                 & 809       & 0.0   & \lrabest{0.02} & 1.15 & 2.67\\
wlan (\texttt{COL}=6, \texttt{MAX\_BACKOFF}=3)
                                          & 284,446   & 0.0   & \lrabest{1.26} & 9.04 & 24.82\\
zeroconf                                  & 1,088     & 0.0   & \lrabest{0.02} & 0.70 & 1.79\\
\bottomrule
\end{tabular}
\end{table}

Our performance comparison is only on MDPs, as this is the model that most tools support.
Thus, we ran \storm (using the default value iteration which provides $\varepsilon$-guarantees), \multigain (using linear programming), and \pet (using value iteration and partial exploration).
We collected benchmarks from several sources~\cite{HKPQR19,AGKM22,BEGKW23};
Appendix G.1 of the full version of~\cite{AGKM22} contains descriptions of many of them.
The experiments were conducted on a freshly installed Ubuntu virtual machine on top of an Intel i7-1165G7 CPU and 8\,GB RAM.
Each run had access to all eight cores available in the virtual machine and the tools were executed sequentially using a bash script, starting with \storm (on all benchmarks) and ending with \pet.
\Cref{tab:mp-perf} shows for every benchmark some characteristics of the model and then the time in seconds taken by each tool (where MG 2.0 is \multigaintwo) to compute the value; the best time is highlighted in bold.
For all but one benchmark, \storm outperforms both \multigain and \pet.
The single exception is mer (4), where \pet is slightly faster, leveraging the fact that only a very small part of the model has to be explored.

\subsubsection{Data availability.}
All model files used for the comparison, as well as the resulting log files, are available at DOI \href{https://doi.org/10.5281/zenodo.8219191}{10.5281/zenodo.8219191}~\cite{LRAArtifact}.

\subsection{Outlook}

We pinpoint several streams of research currently pursued.
First, the algorithms have been extended to \textbf{multiple average rewards}~\cite{BBCFK14,CKK17} and we refer to \Cref{sec:MultiObjective} for a discussion of the achievements and challenges.
Second, some approaches handle \textbf{unknown models} that can only be simulated~\cite{AGKM22,KMMP20}, or avoid their construction for efficiency reasons~\cite{KM20}.
Finally, while value iteration is the prevailing solution approach, \textbf{guarantees on its precision} (stopping criteria) have only been given recently~\cite{ACDKM17,KMW23}.

\section{Linear Temporal Logic}
\label{sec:LTL}

The traditional analysis of MDPs, in particular in the context of operations research and performance optimization, is based on rewards.
In domains such as AI, robotics, or economics, it is often the discounted reward; in other contexts, where the steady state or the long-run behaviour is more relevant, it is the average reward (see \Cref{sec:LRA}).
However, in the context of verification, be it of hardware, software or cyber-physical systems, not only reachability but also more complex temporal properties are required~\cite{Pnu77}.
While the analysis of branching-time properties typically boils down to reachability analysis, the analysis of linear-time properties is more complex.
The most prominent formalism for capturing linear-time properties is the \emph{linear temporal logic} (LTL)~\cite{Pnu77}.

The standard way to analyse LTL properties is the automata-theoretic approach~\cite{VW86}:
The formula is translated to an automaton and, subsequently, the product of the system and this automaton is analysed.
While LTL properties occurring in verification of hardware or non-stochastic software tend to be very complex (see \eg the LTL Store collection~\cite{KMS18a}), this is less pronounced for stochastic systems.
One of the main reasons was the infeasibility of obtaining small automata apt for probabilistic model checking.
Indeed, instead of nondeterministic Büchi automata (NBA), for which good translators had long been available, some degree of determinism is needed for MDP.
Until recently, Rabin automata produced by determinisation were the default but hardly scalable solution.
For instance, about 10 years ago, one fairness contract was translated by the then state-of-the-art methods within \prism to an automaton with 4 states, while a conjunction of two yielded over ten thousand states, and a conjunction of three would not finish computing in a week~\cite{KE12}.

In the decade since~\cite{KE12}, alternative approaches started flourishing.
They avoid the determinisation by direct translations~\cite{EK14} or by employing weaker forms of determinism, such as limit-determinism~\cite{HLSTZ15,SEJK16}.
The resulting tools such as \tool{Rabinizer}~\cite{KMSZ18} or \tool{spot}~\cite{DRCRGSMMDGL22}, or libraries such as \tool{Owl}~\cite{KMS18b}, now reach the same level of scalability as for nondeterministic automata, allowing for verification of more complex properties.
Model checkers such as \prism today
(i)~contain a pre-computed, built-in set of automata for the handful most used properties, and
(ii)~can link external translators to provide the state-of-the-art-sized automata via the Hanoi Omega-Automata (HOA) standard format~\cite{BBDKKMPS15}.
Consequently, comparing the efficiency of model checkers themselves is not very relevant in this category;
instead, we describe the main line of model checking algorithms, and discuss the limitations and additional features of the tools.

\subsection{Algorithms and Tool Support}

The standard algorithm
(i)~constructs the product of the system and the automaton (with a given acceptance condition),
(ii)~identifies the maximum accepting end components, and
(iii)~computes the reachability probability of their union.
Step~(ii) depends on the particular acceptance condition.
The default since the inception of \prism was the Rabin condition due to the better efficiency compared to Muller or parity.
Since the appearance of the new translations, the algorithm has been extended to generalized Rabin~\cite{CGK13} within \prism and limit-determinism in the \prism-based \tool{MoChiBA}~\cite{SK16} as well as in a lazy variant~\cite{HLSTZ15}.
Further improvements on the sizes and types of the automata (\eg good-for-MDP, Emerson-Lei) followed~\cite{EKS20,HPSSTW20,MS17,MBSSZ19,JJBK22}.

Several current tools can be applied to LTL and related specifications:

\begin{description}[itemsep=3pt,topsep=3pt,labelsep=0pt]

\item[\tool{epmc}]
supports the verification of MDPs against LTL and PCTL*.
It translates formulae to an NBA using \tool{spot} and then applies the lazy approach~\cite{HLSTZ15} to compute its satisfaction probability.

\item[\multigaintwo]
is an extension of \prism capable of verification and strategy synthesis for LTL properties combined with long-run average rewards and steady-state constraints.

\item[\prism]
itself supports LTL and PCTL* model checking for MDPs.
LTL model checking is done via deterministic $\omega$-automata, typically (generalised) Rabin automata, simplified to B\"{u}chi or finite automata if appropriate.
The translation uses a custom version of {\tt ltl2dstar}, combined with a built-in automata library; alternatively, it can connect to external translators via the HOA format.
For the subclass of (co)safe LTL, \prism also supports cumulative expected reward until satisfaction properties.

\item[\storm]
answers LTL, lexicographic multi-objective LTL~\cite{CKWW20}, as well as PCTL* queries for MDPs.
It uses deterministic $\omega$-automata with general Boolean acceptance formulas obtained from \tool{spot}.

\end{description}

\subsection{Outlook}

After a decade of research on alternative translations and automata types, the performance both in terms of runtime and of the near-minimality of the size of the automata have reached practical applicability.
A few question remain open, such as whether the semantic notion of \textbf{good-for-MDP automata} allows us to produce yet smaller automata efficiently, compared to the syntactically defined acceptance conditions.
However, the main focus should now move to \textbf{modelling and applications}.
For LTL formulae, a decent amount of benchmarks is available.
Many sets repeatedly used across different papers have been compiled in LTL Store~\cite{KMS18a}.
However due to the earlier scalability problems in probabilistic LTL model checking, the number of probabilistic models coming together with more complex LTL specifications remains quite limited so far~\cite{KNP12,HKPQR19}.

\section{Multi-Objective Analysis}
\label{sec:MultiObjective}

System performance is commonly assessed with respect to multiple quantities such as the probability of a crash, the expected average energy consumption, or the expected time until task completion.
System designers have to consider the interplay between these quantities: minimising the task completion time might require actions that increase the likelihood of a crash.
\emph{Multi-objective analysis}~\cite{CMH06,EKVY08} reveals trade-offs between the considered quantities by showing which 
%
%
\begin{wrapfigure}[11]{r}{0.45\linewidth}
\vspace{-18pt}
\centering
\begin{tikzpicture}
\begin{axis}[
width=1.05\linewidth,
axis on top,
xticklabel style={font=\scriptsize},
yticklabel style={font=\scriptsize},
xlabel style={font=\footnotesize,yshift=2pt},
ylabel style={font=\footnotesize,xshift=-6pt,yshift=-3pt},
xlabel={probability of a crash},
xmin=0,
xmax = 0.1,
scaled x ticks=false,
xticklabel style={/pgf/number format/fixed, /pgf/number format/precision=5},
ylabel={exp. task completion time},
ymin=4,
ymax=12,
axis background/.style={fill=red!30},
axis on top
]
	\addplot[fill=green!30 ] coordinates {
		(0.01, 10 )
		(0.02 , 8 )
		(0.05, 5.5 )
		(0.065, 5 )
		(0.2,5 )
		(0.2,15 )
		(0.01,15 )
	} -- cycle;
	\addplot[blue,ultra thick] coordinates {
		(0.01, 10 )
		(0.02 , 8 )
		(0.05, 5.5 )
		(0.065, 5 )
	};
	\node[circle,  inner sep=1.5pt, fill=black, very thin, label=0:$\vec{p}$] at (axis cs:0.03,9) {};
	\node[circle,  inner sep=1.5pt, fill=black, very thin, label=0:$\vec{q}$] at (axis cs:0.02,8) {};
	\node[align=center,font=\footnotesize] at (axis cs:0.065,10) {achievable\\points};
\end{axis}
\end{tikzpicture}
\end{wrapfigure}
compromises are achievable.
The system is given as a nondeterministic model $\mathcal{M}$ while the quantities are specified as a vector $\langle \varphi_1, \dots, \varphi_\ell \rangle$ of $\ell \ge 2$ objectives.
The objectives commonly refer to rewards attached to states or transitions of $\mathcal{M}$.
An $\ell$-dimensional point $\vec{p} = \langle p_1, \dots, p_\ell \rangle \in \mathbb{R}^\ell$ is \emph{achievable} if there exists a single policy for $\mathcal{M}$---i.e. a resolution of its nondeterminism---under which for all $i \in \{ 1, \dots, \ell\}$ the value of objective $\varphi_i$ is at least (or at most) $p_i$.
Multi-objective analysis answers queries concerning the (set of) achievable points.

As an example, the green area in the figure above on the right shows the set of achievable points for $\ell=2$ objectives as labelled on the axes.
Point $\vec{p} = \langle 0.03,9\rangle$ is achievable but \emph{dominated} by other achievable points; $\vec{q} = \langle 0.02,8\rangle$, for example, yields ``better'' values for both objectives.
The blue line fragments indicate the set of undominated solutions---the \emph{Pareto front}.

\subsection{Algorithms and Tool Support}

\begin{table}[t]
\centering
\caption{Feature comparison of tools for multi-objective verification}
\label{tab:multi:tools}
\setlength{\tabcolsep}{6pt}
\begin{tabular}{lccccc}
\toprule
 & \epmc & \prism & \multigain  & \storm  &  \prismgames \\\midrule\addlinespace[5pt]
models & MDP & MDP  & MDP & MDP, MA & SG \\[2pt]
objectives\\
-- reach. prob. & yes & yes & no & yes  & qualitative\\
-- total reward  & yes  & yes & no   & yes  & yes  \\
-- LRA reward   & no  & no   & yes & yes & yes \\
-- rew. bounded & no & steps  & no& yes &  no   \\
-- LTL prob.   & yes & yes  & yes & lexicographic  & no \\[2pt]
queries \\
-- achievability & yes & yes & yes & yes & yes\\
-- numerical & yes & yes & yes & yes & no \\
-- Pareto & no  & $\ell  =2$  & $\ell  \le 3$ & yes  & $\ell = 2$
\\\bottomrule
\end{tabular}
\end{table}

\Cref{tab:multi:tools} compares quantitative verification tools in terms of their support for multi-objective analysis.
We consider the supported kinds of models (where SG are stochastic games), objectives, and analysis queries.
For the latter, we follow \cite{FKP12} and distinguish
(i)~achievability queries, asking whether a given point $\vec{p}$ is achievable,
(ii)~numerical queries asking for the optimal value for one objective while the others have to achieve a given $(\ell-1)$-dimensional point, and
(iii)~Pareto queries, asking for (an approximation of) the Pareto front.

We elaborate on the features of the individual tools:
\begin{description}[itemsep=3pt,topsep=3pt,labelsep=0pt]

\item[\epmc]
supports multi-objective achievability and numerical queries for MDPs over total reachability reward objectives as well as objectives specified in LTL.
The latter was applied to solve probabilistic preference-based planning problems~\cite{LTHSZ22}.
\epmc implements the algorithm of \cite{FKP12} based on value iteration.

\item[\multigain]
is an extension of \prism that implements the linear pro\-gram\-ming-based approach of~\cite{BBCFK14} for multiple steady-state and LRA reward objectives on MDPs to answer achievability, numerical, and Pareto queries---the latter for up to $\ell = 3$ objectives.
A recent extension \multigaintwo \cite{BEGKW23} also incorporates the methods of~\cite{CKK17,Kre21} to add support for mixtures of steady-state, LRA reward, and LTL specifications.

\item[\prism]
answers achievability, numerical, and Pareto queries for MDPs over combinations of total reward, step-bounded reward, and LTL specification objectives.
It implements methods based on value iteration~\cite{FKP12} and on linear programming~\cite{FKNPQ11}.
While the latter only works for achievability and numerical queries, the former can also be used to approximate Pareto fronts over up to $\ell = 2$ objectives.
\prism's graphical interface allows the user to conveniently examine the results.

\item[\prismgames]
implements value iteration over convex sets~\cite{BKTW15} to analyze multiple total and LRA (ratio) reward objectives as well as almost-sure reachability constraints.
\prismgames supports Pareto queries for $\ell = 2$ objectives and achievability queries for arbitrary Boolean combinations of objectives.
An extension exists towards lexicographic queries for reachability objectives~\cite{CKMWW23}.

\item[\storm]
handles achievability, numerical, and Pareto queries for MDPs and MA~\cite{QJK22}.
\storm implements the algorithm of \cite{FKP12} for total reachability reward objectives as well as extensions towards reward-bounded reachability objectives~\cite{HJKQ20} and LRA reward objectives~\cite{QK21}.
Furthermore, \storm supports multi-dimensional quantile queries~\cite{HJKQ20}, lexicographic LTL specifications~\cite{CKMWW23}, and multi-objective analysis under non-randomised policies with limited memory~\cite{DKQR20}.

\end{description}

\subsection{Performance Comparision}

We empirically compare the performance of the tools for solving achievability queries on MDPs with
(i)~reachability and total reward objectives as well as
(ii)~steady-state and LRA reward objectives.
We consider various benchmark models and objectives from the literature, \eg \cite{KNP12,HKPQR19,FKP12,BCFK15,HJKQ20}.
To obtain challenging achievability queries, the queried points $\vec{p} = \langle p_1, \dots, p_\ell \rangle$ have been obtained by roughly setting the threshold $p_i$ for the $i^{\text{th}}$ objective to $90\%$ of its optimal (single-objective) value.
All experiments ran on an Intel Xeon Platinum 8160 CPU with 8 cores and 32\,GB of RAM available.
We measured the wall-clock runtimes of the tools and aborted executions after 1800 seconds.

\begin{figure}[t]
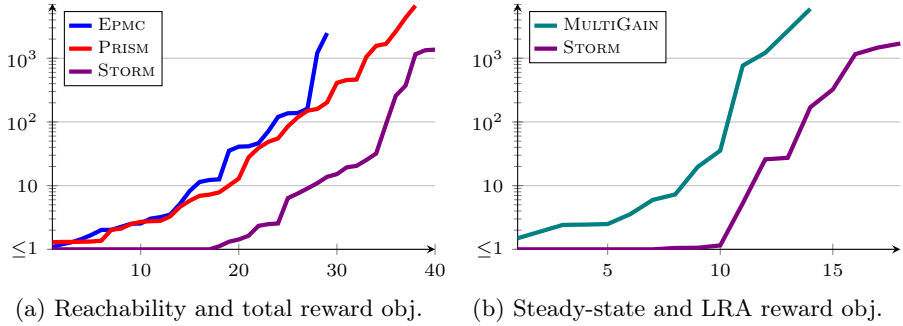

\centering
\begin{minipage}[b]{0.5\textwidth}
\centering
\quantileplot{figures/multi-objective/achtot-quantile.csv}
{epmc.ach/plotblue, prism.ach/plotred, storm.ach/plotpurple}
{\epmc,\prism,\storm}
{1}{40}{1}{7200}{north west}\\
{(a)~Reachability and total reward obj.}
\end{minipage}%
\begin{minipage}[b]{0.5\textwidth}
\centering
\quantileplot{figures/multi-objective/achlra-quantile.csv}
{multigain.ach/plotteal,storm.ach/plotpurple}
{\multigain,\storm}
{1}{18}{1}{7200}{north west}\\
{(b)~Steady-state and LRA reward obj.}
\end{minipage}
\caption{Performance comparison results of tools for multi-objective verification}
\label{fig:multi:results}
\end{figure}

\paragraph{Reachability and total reward objectives.}
Our benchmark set contains 46 concrete queries over reachability probability and total reward objectives from 8 different model families.
These queries are supported by \epmc, \prism, and \storm, which all use the approach of \cite{FKP12} based on value iteration as their default method.
On 12 queries, the tools reported inconsistent achievability results.
We still include these problematic cases in our evaluation since identifying the \emph{correct} solution is not trivial.
The tool runtimes in seconds are summarised in \Cref{fig:multi:results}\,a.
In this \emph{quantile plot}, a point $\langle x , y \rangle$ on a line for a tool means that $x$ out of the 46 queries \emph{each} took at most $y$ seconds to complete with this tool.
We see that \storm is faster than both \prism and \epmc.
The competition among the latter two is tighter, with \prism taking the lead.

\paragraph{Steady-state and LRA reward objectives.}
We consider 20 queries over steady-state and LRA reward objectives from 6 model families. We solve these queries using \multigain and \storm.
All reported results were consistent for this set of experiments.
\Cref{fig:multi:results}\,b summarizes the runtime comparison (again as a quantile plot with runtimes in seconds).
The implementation in \storm using the methods of \cite{QK21} outperforms the linear-programming based approach of \multigain.

\subsubsection{Data availability.}
The benchmark models, scripts to reproduce the experiments, and our tool outputs are available at DOI \href{https://doi.org/10.5281/zenodo.8063883}{10.5281/zenodo.8063883}~\cite{MultiObjectiveArtifact}.

\section{Parametric Markov Models}
\label{sec:Parameters}

Classically, probabilistic model checking assumes that the probabilities on the transitions are fixed and precisely known.
This assumption is often unrealistic:
In various examples, such probabilities are approximations based on expert knowledge.
In other applications, these probabilities reflect design decisions that can be freely made.
\emph{Parametric Markov models} replace constant probabilities by (polynomial) expressions over a fixed set of parameters~$X$.
A~parametric Markov model and a valuation of its parameters induces a standard, parameter-free Markov model.
The analysis of parametric Markov models was introduced almost 20 years ago~\cite{Daw04,LMT07}
while the tool \param brought first tool support more than 10 years ago~\cite{HHZ11}.
By now, the model checkers \storm, \prism and \epmc have support for parametric models.

Over the last decade, there have been various algorithmic advances that answer a variety of \emph{different} queries about a parametric model~\cite{JJK22}.
The accompanying algorithms have been implemented in various tools and prototypes and all make different assumptions.
Furthermore, not every benchmark is well-suited to motivate a particular query.
This situation harms further adoption.
In the spirit of QComp, we provide a unified and cleaned-up reference implementation on top of the probabilistic model checker \storm, and an annotated benchmark set for various parametric verification queries on parametric DTMCS (pDTMCs) and parametric MDPs (pMDPs).

\subsection{Queries and Algorithms}

We formulate the key verification tasks for parametric Markov chains.
For conciseness, we assume that we are interested in computing the expected reward until reaching a target state, which generalizes computing reachability probabilities.
For pMDPs, we assume that we consider the maximal expected reward.
The key queries we identify are as follows:

\begin{description}[itemsep=3pt,topsep=3pt]

\item[Feasibility:]
Find parameter values such that the induced expected reward is above a threshold.
The state-of-the-art methods rely on guess-and-verify and guess using sampling~\cite{CHHKQZ13}, gradient descent~\cite{HSJMK22}, and convex optimisation~\cite{CJJKT22}; the former methods are fastest with few parameters and the latter are fastest with a larger number of parameters.
For pMDPs, the quantification order is first over the parameters and then over the schedulers, \ie the scheduler may depend on the parameter value chosen.
This contrasts with \emph{robust} schedulers that do not allow this~\cite{ABCKS18,WJPK19}.

\item[Verification:]
Show that no parameter values exist such that the induced expected reward is above a threshold.
This problem is the dual to feasibility queries, but the universal quantification makes it harder to solve.
The state-of-the-art approach employs an abstraction-refinement loop~\cite{JAHJKQV19} using interval Markov chains and combines this with a graph-based analysis to determine monotonicity of the parameters~\cite{SJK21}.

\item[Solution function computation:]
Compute a function that maps parameters to the induced expected reward in the corresponding Markov model.
While various dedicated methods exist~\cite{HHZ11,FTG16,JAHJKQV19}, linear equation solving over the field of rational functions performs great overall~\cite{JAHJKQV19}.
Theoretically, a one-step fraction-free method prevents intermediate blowups~\cite{BHHJKK20}.
For pMDPs, the shapes of these functions are typically prohibitive.

\end{description}
A formal treatment is provided in~\cite{Jun20}.
Various other queries have been discussed in the literature.
They aim to partition the parameter space~\cite{JAHJKQV19}, repair a model with the best parameter values~\cite{BGKRS11}, quickly sample parametric Markov models~\cite{GHS18}, or check whether the derivative is (globally) positive~\cite{SJK19}.
Others assume a distribution over parameter values~\cite{BS18,BCJJKT22}.

\paragraph{Practicalities.}
Typically, approaches limit the type of parameter valuations that are considered;
graph-preserving valuations require that the underlying graph does not change.
While this does not change the complexity of \eg feasibility~\cite{WJPK19}, it means that the solution function for pDTMCs is a (continuous) rational function and simplifies preprocessing.
Likewise, most approaches assume that all pDTMCs are \emph{simple(x)}~\cite{Jun20}, which (roughly) means that the transition probabilities are given by affine functions that syntactically sum to one.
While it is theoretically relevant to allow real-valued parameter valuations, tools typically restrict themselves to rational (or floating-point) number representations.

\subsection{Benchmark Collection}

We provide a benchmark collection with 12 benchmark families at
$$
\text{\href{https://github.com/sjunges/parametric-Markov-models}{github.com/sjunges/parametric-Markov-models}}
$$
(with the models also archived at DOI \href{https://doi.org/10.5281/zenodo.10646479}{10.5281/zenodo.10646479}~\cite{ParametersArtifact}).
This benchmark collection includes reference invocations for \storm.
The collection includes parametric versions of classical benchmarks~\cite{KNP12,HKPQR19,HHZ11} as well as benchmarks based on the usage of parametric verification in the analysis of hierarchical MDPs~\cite{JS22} and from the sensitivity analysis of Bayesian networks~\cite{SK21}.

\subsection{Outlook}

We believe that the engineering behind many algorithms is still naive and that there is great potential for \textbf{algorithmic improvements}.
In particular, the verification algorithms lack severely behind in their scalability, and despite being built on top of probabilistic model checking engines, most algorithms have only been implemented for expected rewards and reachability probabilities.
More fundamentally, the \textbf{synthesis of robust policies in pMDPs} is an open challenge.
A next iteration of QComp could also include parametric CTMCs~\cite{BCDS13}, parametric PTAs~\cite{HKKS21}, or structural parameters~\cite{ACJKS21}.

\section{Partially-Observable MDPs}
\label{sec:POMDPs}

A major shortcoming of the classic MDP framework is the assumption that decisions can be made based on \emph{complete} state information.
In many domains where reasoning about uncertainty is necessary, this assumption is not realistic.
For example, information about the current state of an autonomous vehicle is inherently incomplete as it perceives its environment through imperfect sensors.

\emph{Partially observable MDPs} (POMDPs) extend MDPs with the notion of \emph{partial observability}.
Nondeterminism is resolved not based on the complete state, but on the observable information available to the decision procedure.
As such, policies for POMDPs are required to be \emph{observation-based}, \ie decisions are based on the observations and their history.
We consider reachability objectives in POMDPs, \ie we are interested in the minimal or maximal \emph{reachability probability} of certain states in the system or, alternatively, in the minimal and maximal expected total reward until reaching a set of states.
In contrast to fully observable MDPs, where optimal policies for such objectives that are memoryless always exist, in POMDPs memory is crucial even for sub-optimal policies.

While POMDPs are widely used for planning in domains like artificial intelligence~\cite{RN20}, many verification and synthesis problems have proven to be undecidable.
For example, even determining if the reachability probability of a set of states in a POMDP exceeds a threshold is undecidable~\cite{MHC03}.
Tool support for POMDPs exists in the AI community~\cite{KHL08,ESBWGK17}, however, these tools focus on \emph{discounted} objectives, often over a finite horizon~\cite{SPK13}.
In recent years, efforts took place to extend the tool support for the verification setting of infinite-horizon objectives \emph{without} discounting, also called \emph{indefinite-horizon} objectives.
Due to the undecidability of key problems in this setting, the applied methods focus on approximating values and synthesising \emph{good} (sub-optimal) policies with respect to the objectives.

\subsection{Algorithms and Tool Support}
\label{sec:pomdp_tool_overview}

We showcase three tools from the formal methods community that deal with verification and policy synthesis for POMDPs.
\begin{description}[itemsep=3pt,topsep=3pt,labelsep=0pt]

\item[\prism]
includes support for POMDPs as well as a partially observable variant of PTA.
It solves probabilistic reachability or expected cumulative reward queries using the model checking algorithm of~\cite{NPZ17}, which implements a grid-based approach~\cite{Lov91,YB04} for computing an over-approximation of the objective value on an abstraction of the infinite, fully observable \emph{belief MDP} of the POMDP using only a finite number of beliefs.
The resulting policy is then solved to yield an under-approximation which, together, provides lower and upper bounds on the objective value for the POMDP.
If the bounds are not tight enough, the approximation can be refined by increasing the grid resolution.
The implementation builds upon \prism's Java-based explicit-state engine.
The tool then allows the resulting policy to be visualised or simulated in its graphical user interface.

\item[\storm]
has support for POMDPs that is actively in development.
In contrast to \prism, over- and under-approximations can be computed independently of each other.
For over-approximations, \storm implements an improved version of the grid-based approach from~\cite{Lov91}:
it allows for variable grid resolutions for different observations and on-the-fly generation of the belief grid~\cite{BJKQ20}.
For under-approximations, \storm uses \emph{belief unfolding with cut-offs}:
the belief MDP is unfolded starting in the initial belief.
After a fixed number of beliefs has been unfolded, the objective value for all beliefs that have not yet been fully expanded are approximated.
This approximation is based on values computed on the POMDP itself using \emph{some} observation-based policy~\cite{BKQ22}.
These values can be computed heuristically by \storm or provided externally.
The abstract MDPs are then checked using \storm's MDP model checking core.
In addition to the quantitative properties considered here, \storm also supports the verification of \emph{qualitative} properties on POMDPs~\cite{JJS21}.

\item[\paynt]
was originally developed for the inductive synthesis of probabilistic programs.
In contrast to the model checkers described above, it aims at directly synthesising \emph{finite-state controllers} (FSCs) for POMDPs.
An FSC is a Mealy machine that compactly represents a finite-memory policy.
To find the best FSC within a given design space of controllers, an MDP abstraction is used, which encodes every possible decision and memory update a policy can make.
The resulting process is an over-approximation in the sense that it can simulate every FSC in the design space and switch between FSCs mid-execution.
Model checking the MDP yields the best policy within the design space and, if the policy is not observation-based, a refinement takes place.
Additionally, the design space can be pruned by generating counter-examples~\cite{ACJK22}.
For computations on the MDP abstraction and assessing the quality of a synthesised FSC, \paynt internally uses \storm.
As \paynt synthesises a policy, it can only provide under-approximations of the objective values.

\end{description}

\subsection{Performance Comparison}

We empirically evaluate the tools described above.
As the different approaches and features of the tools make a direct comparison between them misleading, we consider the tools separately on different benchmarks.
All three tools accept as input descriptions of POMDPs in the \prism format.
Furthermore, \storm and \paynt allow inputs in the explicit DRN format.

From the repository of POMDP benchmarks from the literature~\cite{NPZ17,BKQ22,ACJK22}~at
$$
\text{\href{https://github.com/moves-rwth/pomdp-collection}{github.com/moves-rwth/pomdp-collection},}
$$
we select one benchmark for each tool to showcase some of its capabilities.
All experiments ran on an Intel Xeon Platinum 8160 CPU using 2 threads, 32\,GB of RAM, and a time limit of 1\,h (measured in wall time).
All tools are called using default configurations and options except for the input parameter we evaluate.
The short evaluation presented here is not representative of the full capabilities of the~tools. Tweaking the configurations used for running the tools typically leads to improvements in the results obtained.

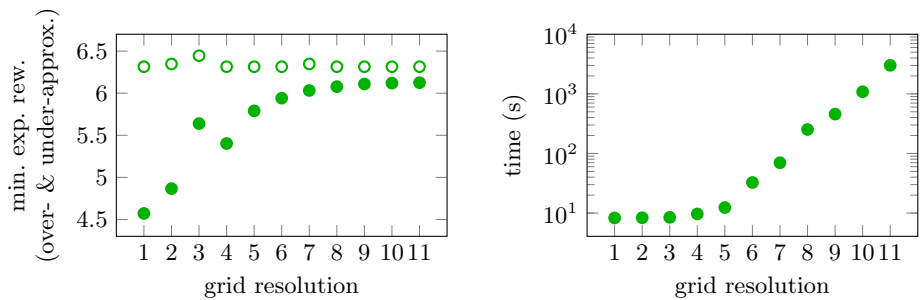
\begin{figure}[tp]
\centering
\begin{tikzpicture}
    \begin{axis}[width=0.49\textwidth,
			    height=0.35\textwidth,
 				xlabel={grid resolution},
 				xmin=0,
 				xmax = 12,
 			 	xtick={1,2,3,4,5,6,7,8,9,10,11},
 				xticklabel style={/pgf/number format/fixed, /pgf/number format/precision=5},
 				scaled x ticks=false,
 				ylabel={min.\ exp.\ rew.\\ (over- \& under-approx.)},
                ylabel style={align=center},
 				ymin=4.3,
 				ymax=6.7,
 				ytick={4.5,5,5.5,6,6.5},
 				axis on top,
 				legend style={at={(0.95,0.15)}, anchor=east, legend columns=-1}]
        \addplot[mark=*, mark size=2pt, black!30!green, thick, only marks ] table [x=x, y=y, col sep=comma] {figures/pomdps/prism_over.csv};
        \addplot[mark=o, mark size=2pt, black!30!green, thick, only marks ] table [x=x, y=y, col sep=comma] {figures/pomdps/prism_under.csv};
    \end{axis}
\end{tikzpicture}
\hfill
\begin{tikzpicture}
    \begin{axis}[width=0.49\textwidth,
			    height=0.35\textwidth,
 				xlabel={grid resolution},
 				xmin=0,
 				xmax = 12,
 			 	xtick={1,2,3,4,5,6,7,8,9,10,11},
 				xticklabel style={/pgf/number format/fixed, /pgf/number format/precision=5},
 				scaled x ticks=false,
 				ylabel={time (s)},
 				ymax=10000,
 				ytick={10,100,1000,10000},
                ymode=log,
 				axis on top,
 				legend style={at={(0.95,0.15)}, anchor=east, legend columns=-1}]
        \addplot[mark=*, mark size=2pt, black!30!green, thick, only marks ] table [x=x, y=y, col sep=comma] {figures/pomdps/prism_times.csv};
    \end{axis}
\end{tikzpicture}
\caption{\prism's over- and under-approximations for \emph{grid} and computation times}
\label{fig:prism}
\end{figure}

\paragraph{\prism.}
We consider the \emph{grid} benchmark, instantiated with length 4 and slipping probability $0.3$, for our evaluation of \prism.
The objective here is to minimise an expected reward.
We evaluate \prism with respect to changes in the resolution of the belief grid considered for the over-approximation.
Our results are depicted in \Cref{fig:prism}.
The grid-based over-approximation---in the case of minimisation a lower bound---is depicted with solid dots while the under-approximation is depicted with circles.
The experiments clearly show the impact of increasing the grid resolution.
Generally, the higher the resolution, the better the computed bounds, for the over-approximation as well as the under-approximation which is only indirectly effected by the grid resolution.
However, this improvement comes at the cost of greatly increased runtimes.
The outlier at resolution $3$ also shows that a good choice of resolution may lead to a tighter approximation of the actual belief space even for rather small resolutions.
For resolutions greater than $11$, the tool either times out or runs out of memory.

\begin{figure}[tp]
\centering
\begin{tikzpicture}
    \begin{axis}[width=0.49\textwidth,
			    height=0.35\textwidth,
 				xlabel={grid resolution},
 				xmin=0,
 				xmax = 6,
 			 	xtick={1,2,3,4,5},
 				xticklabel style={/pgf/number format/fixed, /pgf/number format/precision=5},
 				scaled x ticks=false,
 				ylabel={max.\ prob.\\ (over-approx.)},
                ylabel style={align=center},
 				ymin=0.45,
 				ymax=1.1,
 				ytick={0.5,0.6,0.7,0.8,0.9,1},
 				axis on top,
 				legend style={at={(0.95,0.15)}, anchor=east, legend columns=-1}]
        \addplot[mark=triangle, mark size=2pt, blue, thick, only marks ] table [x=x, y=y, col sep=comma] {figures/pomdps/storm_over_result.csv};
    \end{axis}
\end{tikzpicture}
\hfill
\begin{tikzpicture}
    \begin{axis}[width=0.49\textwidth,
			    height=0.35\textwidth,
 				xlabel={grid resolution},
 				xmin=0,
 				xmax = 6,
 			 	xtick={1,2,3,4,5},
 				xticklabel style={/pgf/number format/fixed, /pgf/number format/precision=5},
 				scaled x ticks=false,
 				ylabel={time (s)},
 				ytick={0.1,1,10,100,1000},
                ymode=log,
 				axis on top,
 				legend style={at={(0.95,0.15)}, anchor=east, legend columns=-1}]
        \addplot[mark=triangle, mark size=2pt, blue, thick, only marks ] table [x=x, y=y, col sep=comma] {figures/pomdps/storm_over_time.csv};
    \end{axis}
\end{tikzpicture}
\caption{\storm's over-approximations for \emph{refuel} and computation times}
\label{fig:storm_over}
\end{figure}
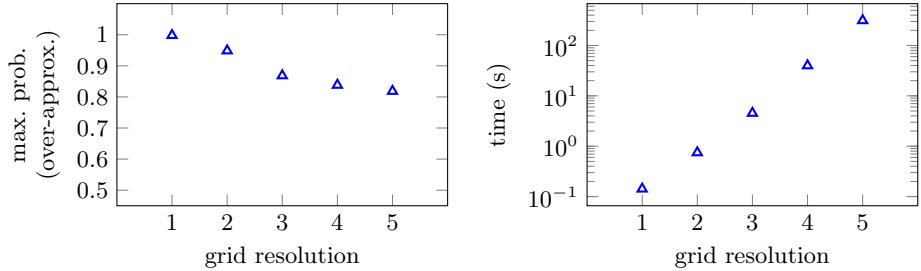

\begin{figure}[tp]
\centering
\begin{tikzpicture}
    \begin{axis}[width=0.49\textwidth,
			    height=0.35\textwidth,
 				xlabel={exploration size},
 				xmin=0,
 			 	xtick={1000000,3000000,5000000},
 				scaled x ticks=false,
 				ylabel={max.\ prob.\\ (under-approx.)},
                ylabel style={align=center},
 				axis on top,
 				legend style={at={(0.95,0.15)}, anchor=east, legend columns=-1}]
        \addplot[mark=triangle, mark size=2pt, blue, thick, only marks ] table [x=x, y=y, col sep=comma] {figures/pomdps/storm_under_result.csv};
    \end{axis}
\end{tikzpicture}
\hfill
\begin{tikzpicture}
    \begin{axis}[width=0.49\textwidth,
			    height=0.35\textwidth,
 				xlabel={exploration size},
 				xmin=0,
 			 	xtick={1000000,3000000,5000000},
 				scaled x ticks=false,
 				ylabel={time (s)},
                ymode=log,
 				axis on top,
 				legend style={at={(0.95,0.15)}, anchor=east, legend columns=-1}]
        \addplot[mark=triangle, mark size=2pt, blue, thick, only marks ] table [x=x, y=y, col sep=comma] {figures/pomdps/storm_under_time.csv};
    \end{axis}
\end{tikzpicture}
\caption{\storm's under-approximations for \emph{refuel} and computation times}
\label{fig:storm_under}
\end{figure}
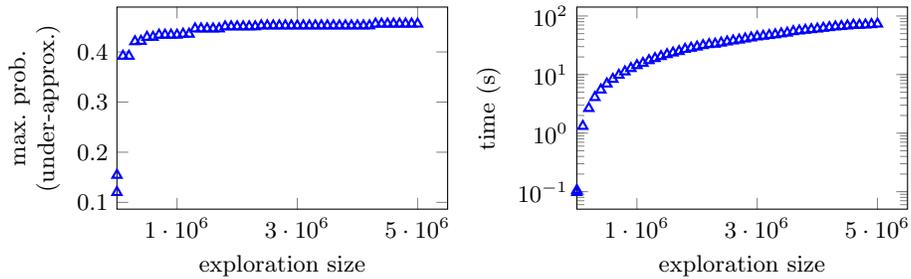

\paragraph{\storm.}
We evaluate \storm on the \emph{refuel} benchmark instantiated with length 10, computing the maximal probability.
We study the over-approximation with respect to changes in the grid resolution and the under-approximation with respect to the size of the unfolding of the belief MDP.
\Cref{fig:storm_over,fig:storm_under} depict the respective results.
Like for the over-approximation in \prism, we observe that an increase in resolution for the grid-based approach leads to tighter bounds on the optimal objective value, with an impact on the runtime.
For higher resolutions than $5$, \storm ran out of memory on the benchmark.
For the under-approximation, we consider unfoldings with increasingly larger state spaces.
For larger unfoldings than depicted, we did not achieve better values.
We see that with increasing unfolding depth, a tighter bound is achieved.
This effect is particularly pronounced in the range of smaller unfolding sizes.
With a linear increase in the unfolding size, runtimes appear to scale proportionally.

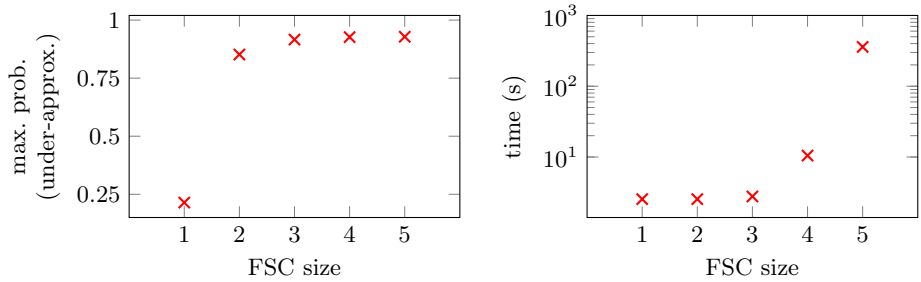
\begin{figure}[tp]
\centering
\begin{tikzpicture}
    \begin{axis}[width=0.49\textwidth,
			    height=0.35\textwidth,
 				xlabel={FSC size},
 				xmin=0,
 				xmax = 6,
 			 	xtick={1,2,3,4,5},
 				xticklabel style={/pgf/number format/fixed, /pgf/number format/precision=5},
 				scaled x ticks=false,
 				ylabel={max.\ prob.\\ (under-approx.)},
                ylabel style={align=center},
 				ymin=0.15,
 				ymax=1.02,
 				ytick={0.25,0.5,0.75,1},
 				axis on top,
 				legend style={at={(0.95,0.15)}, anchor=east, legend columns=-1}]
        \addplot[mark=x, mark size=3pt, red, thick, only marks ] table [x=x, y=y, col sep=comma] {figures/pomdps/paynt_results.csv};
    \end{axis}
\end{tikzpicture}
\hfill
\begin{tikzpicture}
    \begin{axis}[width=0.49\textwidth,
			    height=0.35\textwidth,
 				xlabel={FSC size},
 				xmin=0,
 				xmax = 6,
 			 	xtick={1,2,3,4,5},
 				xticklabel style={/pgf/number format/fixed, /pgf/number format/precision=5},
 				scaled x ticks=false,
 				ylabel={time (s)},
 				ymax=1000,
 				ytick={10,100,1000},
                ymode=log,
 				axis on top,
 				legend style={at={(0.95,0.15)}, anchor=east, legend columns=-1}]
        \addplot[mark=x, mark size=3pt, red, thick, only marks ] table [x=x, y=y, col sep=comma] {figures/pomdps/paynt_times.csv};
    \end{axis}
\end{tikzpicture}
\caption{Max.\ prob.\ achieved and time to compute the FSC for \emph{grid-avoid} with \paynt}
\label{fig:paynt}
\end{figure}

\paragraph{\paynt.}
For evaluating \paynt, we select the \emph{grid-avoid} benchmark, instantiated with length $4$ and slipping probability $0.1$, where the objective is to maximise the probability to reach a target.
While a key feature of \paynt is its ability to dynamically increase the size of the considered FSC during computation, we focus on its functionality to compute FSCs of a given size.
Thus, we vary the corresponding input parameter.
Our results in \Cref{fig:paynt} show that \paynt is able to obtain FSCs using \emph{some} memory very quickly while FSC quality greatly improves when memory is considered.
With increasing FSC size, however, the effect is far less pronounced while runtimes increase drastically.
For all FSC sizes greater than $5$, the tool timed out in our experiment.

\subsubsection{Data availability.}
The benchmark models used in and the tool outputs generated by our experiments are available at DOI \href{https://doi.org/10.5281/zenodo.8215337}{10.5281/zenodo.8215337}~\cite{PomdpsArtifact}.

\section{Rare Events}
\label{sec:RareEvents}

In stochastic systems, \emph{rare events} (\RE) stand for measurable events with a positive but very low probability of occurrence.
A typical example is the failure probability of highly-reliable systems that can be from $10^{-3}$ to $10^{-20}$ or lower.

Formal methods tools can encounter \RE in quantitative computations of PRCTL- or CSL-like queries on any model with probabilities.
Queries on general stochastic models---\ie models in continuous-time with residence times or transition firings governed by arbitrary probability distributions---are often estimated via Monte Carlo simulation.
This is known as \emph{statistical model checking} and is hindered by \RE: since the states of interest are seldom visited, either the number of simulation runs required and thus the runtime grows to impractical values, or the statistical estimations become imprecise or even incorrect.
The field of \emph{rare event simulation} (\RES) has developed to tackle this problem, and can be divided in two main approaches:
\emph{importance sampling} (\IS~\cite{Hei93})
and
\emph{importance splitting} (\ISPLIT~\cite{KH51}).

In contrast, exhaustive state-space exploration approaches such as (probabilistic) model checking are not hindered by the rarity of an event, which makes them very attractive to solve \RE property queries.
However, model checking struggles with the state-space explosion problem, which can be portrayed as the counterpart of the runtime explosion problem faced by simulation analyses.
Additionally, no scalable exhaustive methods are available for models with general distributions.
Thus, the challenge here---in the Markovian case---is how to reduce the model size without compromising the correctness or accuracy of the final estimate.

We compare tools that implement \RES and model checking to estimate quantitative \RE queries in formal stochastic models.
Besides defining the scope and capabilities of each tool, we showcase their computation of \RE queries in six models with Markovian and arbitrary probability distributions.

\subsection{Algorithms and Tool Support}
\label{sec:tools}

\begin{table}[t]
	\centering
	\caption{Feature comparison of tools for rare event estimation}
	\label{tab:tools}
	\smaller
	\setlength{\tabcolsep}{2pt}
	\renewcommand{\arraystretch}{1.1}
	\def\SHIFTL{\hspace*{-1em}}
	\begin{tabular}{
		>{}   L{5.5em} @{~\:}   	
		>{}   L{6.6em} @{~~} 	
		>{}   C{6.8em} @{~}   	
		>{}   C{6.2em} @{~~}   	
		>{}   C{2.8em} @{~~}   	
		>{}   C{5.4em} @{~~}    	
		>{}   C{5.9em} @{\;} 	
	}
	\toprule
	\multicolumn{2}{>{\SHIFTL}c@{\quad}}{Tool}
	& \multicolumn{2}{>{\SHIFTL}c@{~~}}{Approach to RE}
	& \multicolumn{2}{>{\SHIFTL}c@{~~}}{Models}
	& 
	\\
	\cmidrule(l{.7\tabcolsep}r{2.5\tabcolsep}){1-2}
	\cmidrule(l{.7\tabcolsep}r{2.5\tabcolsep}){3-4}
	\cmidrule(l{.7\tabcolsep}r{2.5\tabcolsep}){5-6}
	Name & OS
	& Type & Subtypes
	& Types & Syntax
	& \multirow{-2}{6em}{\centering Semantic formalism}
	\\
	\midrule
	\dftres & Linux, macOS, Windows
	& \RES: \IS & Path-ZVA, forcing
	& DFT, RFT & \jani, Galileo
	& CTMC, MA~(subset)
	\\[12pt]
	\fig & Linux
	& \RES: \ISPLIT & \RESTARTPj, fixed effort
	& IOSA, DFT, RFT & IOSA, \jani, Galileo, Kepler
	& CTMC, IOSA
	\\[20pt]
	\modes & Linux, macOS, Windows
	& \RES: \ISPLIT & \RESTARTPj, fixed effort
	& any & Modest, \jani
	& DTMC, CTMC, SHA
	\\[12pt]
	\ragtimer & Linux
	& probabilistic model checking & partial exploration
	& CRN & \ragtimer
	& CTMC
	\\[12pt]
	\tool{St.DftRes} & Linux, macOS
	& \RES: \ISPLIT & \RESTART
	& DFT & Galileo
	& CTMC
	\\
	\bottomrule
	\end{tabular}
\end{table}

\Cref{tab:tools} summarises the characteristics of tools in formal methods that can estimate rare events.
Naturally, the list is not exhaustive: see \eg earlier works \cite{BHP12,JLS16,MNBDLB18}.
In QComp 2023, we compare the following tools for rare event scenarios, of which all except \ragtimer are statistical model checkers:
\begin{description}[itemsep=3pt,topsep=3pt,labelsep=0pt]

\item[\dftres]
is designed to estimate transient and steady-state properties of repairable DFTs specified in Galileo, and can also be applied to more general CTMCs and some MAs (that reduce to CTMCs) specified in \jani.
It automatically applies \IS \RES, through the Path-ZVA algorithm~\cite{RBSJ18} and, for transient properties, using forcing~\cite{NSN01}.
These techniques are particularly applicable for models in which the target event can be reached in a relatively small number of low-probability steps.
The algorithms can be applied with no user adjustments, however manual tweaking can improve performance on specific models.

\item[\fig]
estimates transient and steady state reachability properties in CTMCs and IOSA.
It can parse the IOSA and \jani syntax for general models, and Galileo and Kepler for repairable DFTs.
\fig automates the use of \ISPLIT \RES by deriving the importance function from the model and property query~\cite{BDM16}.
This simplifies the user input to choosing a thresholds-selection algorithm (sequential Monte Carlo or expected success~\cite{BDH19}), a simulation run type (\RESTART, \RESTARTPj, or fixed effort), and termination criteria (\eg by runtime length).
There are no theoretical proofs---\eg of asymptotic efficiency---on the convergence time of the algorithms used by the tool on general models.

\item[\modes]
implements \ISPLIT to tackle \RES with manually-specified or automati\-cally-derived importance functions much like \fig, including support for the same run types and thresholds-selection algorithms~\cite{BDH19,BH21}.

\item[\ragtimer]
uses guided stochastic simulation and commutability properties to build a partial state space and acquire a lower-bound for a rare-event probability in chemical reaction networks (CRNs) modeled as CTMCs.
It uses reaction information to create a dependency graph, which can demonstrate unreachability.
If a property is reachable, it constructs a prob\-ab\-il\-ity-\-agnostic model for compositional testing in the IVy tool~\cite{MZ19} and uses stochastic simulation to generate a large number of counterexample traces.
It then expands these traces and discovers parallel traces by firing commutable reactions and cycles from every state along a trace.
The resulting partial state space is passed explicitly to a probabilistic model checker to obtain a lower-bound on the probability of interest.
In preliminary testing, \ragtimer finds or approaches the true probability of rare-event properties in several CRN models.

\item[\stormdftres]
analyses time-bounded reachability properties on (non-repair\-able) DFTs in either the Galileo or a custom \acronym{json} format represented as CTMCs.
It aims to perform importance splitting for \RE following the ideas of \fig and using the importance functions for DFTs presented in~\cite{BDMS22}.

\end{description}

\subsection{Performance Comparison}
\label{sec:experiments}

\label{sec:experiments:models}

\begin{table}[t]
	\centering
	\caption{Models used in the comparison of tools for rare event estimation}
	\label{tab:models}
	\def\squeezedCTMC{\parbox[t]{\linewidth}{\centering%
		\acronym{c\hspace{.5pt}t}\\[-.7ex]\acronym{mc}}}  
	\smaller
	\setlength{\tabcolsep}{2pt}
	\renewcommand{\arraystretch}{1.1}
	\begin{tabular}{
		>{}   L{4.0em}  @{}  	
		>{}   C{2.5em}  @{~}   	
		>{}   L{4.8em}  @{\,}  	
		>{}   L{17.7em} @{~} 	
		>{}   L{12.5em} @{~~}  	
	}
	\toprule
	Name
	& \hspace{-5pt} Type
	& Family
	& Description
	& Properties
	\\ \midrule
	forked- ${}$~cycle- ${}$~\mbox{tandem}- ${}$~queue
	& \squeezedCTMC
	& queueing system
	& \parbox[t]{\linewidth}{%
    \emph{three queues}:
    arrivals to Q1;
    probabilistic output to Q1, Q2, Q3;
    study overflows in Q2.\\
    \hfill \textsl{\color{black!70}(previously unpublished)}~
    }
	& \parbox[t]{\linewidth}{%
		\Prop[1]: \PropP{ q2 > 0 \PropU q2 \geqslant 6}\\
        \Prop[2]:~\PropP{ q2 > 0 \PropU[\leqslant555] q2 \geqslant 6}\\
        \Prop[3]: \PropP{ \PropF[\leqslant555] (q2 \geqslant 6)}\\
        \Srop[4]: \PropS{q2 \geqslant 6}
	}
	\\[30pt]
	7nodes- ${}$~network
	& SA
	& queueing system
	& \emph{non-Jackson 7 queues}:
    arrivals to all queues; near-full probabilistic interconnection; study overflow in Q7.
    \hfill\cite{VA18}~~
	& \Srop[5]: \PropS{n7 \geqslant 30}
	\\[20pt]
	2react
	& \squeezedCTMC
	& chemical reaction network
	& \emph{single species production-degradation}:
    simple 2-reaction system with one shortest trace.
    \hfill\cite{DRGP11}~~
	& \Prop[6]: \PropP{\PropF[\leqslant100] (s2 \geqslant 80) }
	\\[20pt]
	6react
	& \squeezedCTMC
	& chemical reaction network
	& \emph{enzymatic futile cycle}:
    6-reaction system with large state space, cyclic behavior, and one shortest trace.
    \hfill\cite{KM08}~~
	& \Prop[7]: \PropP{\PropF[\leqslant100] (s5 = 25) }
	\\[20pt]
	8react
	& \squeezedCTMC
	& chemical reaction network
	& \emph{modified yeast polarization}:
    concurrent 8-reaction system with cyclic behavior and many shortest traces.
    \hfill\cite{DSFZ13}~~
	& \Prop[8]: \PropP{\PropF[\leqslant20] (G\_bg \geqslant 50) }
	\\[20pt]
	\acronym{hecs}
	& \squeezedCTMC
	& dynamic fault tree
	& \emph{hypothetical example computer system}:
    standard DFT benchmark.
    \hfill\cite{VJK16}~~
	& \parbox[t]{\linewidth}{%
        \Prop[9~]: \PropP{\PropF[\leqslant1] TLE}\\
		\hfill``\textit{unreliability @ 1}''
    }
	\\[14pt]
	\acronym{mas}
	& \squeezedCTMC
	& dynamic fault tree
	& \emph{mission avionics system}:
    highly redundant safety-critical system with hard- and software components.
    \hfill\cite{PD02}~~
	& \parbox[t]{\linewidth}{%
        \Prop[10]: \PropP{\PropF[\leqslant1] TLE}\\[.5ex]
		\hfill``\textit{unreliability @ 1}''
    }
	\\[17pt]
	\bottomrule
	\end{tabular}
\end{table}

We demonstrate the capabilities of the tools to compute various PRCTL- and CSL-like property queries on a series of CTMC and SA models.
The models used for experimentation are summarised in \Cref{tab:models}:
there are six Markovian (CTMCs) and one non-Markovian (SA) models, the latter with hyperexponential and Erlang distributions.
All models are provided in the syntax of the tool that specialises in it, and which introduced it to this comparison.
They have also been translated to \jani, for the model exchange across tools that enabled this comparison.
The SA model is an exception: it is provided in the IOSA and \lang{Modest} syntaxes alone (for the \fig and \modes tools), since it has committed actions that are currently unsupported in \jani.

\label{sec:experiments:results}

\begin{table}[tp]
  \centering
  \caption{Performance comparison results of tools for rare event estimation}
  \label{tab:experiments}
  \def\cuscmd#1{{#1}\textsuperscript{$\star$}\!\!\!}
  \def\error{\ensuremath{\varnothing}\xspace}
  \def\ditto{\raisebox{.7ex}{\normalsize\,%
    \rotatebox{90}{-}\hspace{-2pt}\rotatebox{90}{-}}}
  \smaller
  \colorlet{shade}{gray!21}
  \setlength{\tabcolsep}{12pt}
  \renewcommand{\arraystretch}{1.1}
  \begin{tabular}{
		>{}  c       @{~~}   	
		>{}  c       @{\quad} 	
		>{}  c       @{\quad} 	
		>{}  c       @{~~~}   	
		>{}  c       @{~\quad}  
		>{}  c       @{~~}    	
		>{}  c       @{~~}    	
		>{}  c       @{}    	
	}
	\toprule
	\multicolumn{2}{l}{Prop.}
	& \raisebox{-0.5pt}{\larger\clock}
	& \bfseries \dftres
	& \bfseries \fig
	& \bfseries \modes
	& \bfseries \ragtimer
	& \bfseries \stormdftres
	\\
	\midrule
	& 
	& 1
	& $9.2\eminus{10} \pm 3\eminus{13}$ 
	& \cuscmd{$9.0\eminus{10} \pm 9\eminus{11}$} 
	& \cuscmd{$9.4\eminus{10} \pm 6\eminus{11}$} 
	& 
	& 
	\\
	& 
	& 5
	& $9.2\eminus{10} \pm 6\eminus{14}$ 
	& \cuscmd{$9.5\eminus{10} \pm 4\eminus{11}$} 
	& \cuscmd{$9.2\eminus{10} \pm 4\eminus{11}$} 
	& 
	& 
	\\
	& 
	& 10
	& $9.2\eminus{10} \pm 4\eminus{14}$ 
	& \cuscmd{$9.0\eminus{10} \pm 3\eminus{11}$} 
	& \cuscmd{$9.2\eminus{10} \pm 2\eminus{11}$} 
	& 
	& 
	\\
	\multirow{-4}{*}{\Prop[1]}
	& \multirow{-4}{*}{\rotatebox{90}{9.23\eminus{10}}}
	& 30
	& $9.2\eminus{10} \pm 2\eminus{14}$ 
	& \cuscmd{$9.3\eminus{10} \pm 2\eminus{11}$} 
	& \cuscmd{$9.1\eminus{10} \pm 8\eminus{11}$} 
	& 
	& 
	\\[4pt]
	& 
	& 1
	& $9.2\eminus{10} \pm 5\eminus{13}$ 
	& \cuscmd{$9.4\eminus{10} \pm 1\eminus{10}$} 
	& \cuscmd{$9.3\eminus{10} \pm 3\eminus{10}$} 
	& 
	& 
	\\
	& 
	& 5
	& $9.2\eminus{10} \pm 8\eminus{14}$ 
	& \cuscmd{$9.2\eminus{10} \pm 5\eminus{11}$} 
	& \cuscmd{$9.1\eminus{10} \pm 3\eminus{11}$} 
	& 
	& 
	\\
	& 
	& 10
	& $9.2\eminus{10} \pm 6\eminus{14}$ 
	& \cuscmd{$9.3\eminus{10} \pm 3\eminus{11}$} 
	& \cuscmd{$9.2\eminus{10} \pm 2\eminus{11}$} 
	& 
	& 
	\\
	\multirow{-4}{*}{\Prop[2]}
	& \multirow{-4}{*}{\rotatebox{90}{9.23\eminus{10}}}
	& 30
	& $9.2\eminus{10} \pm 3\eminus{14}$ 
	& \cuscmd{$9.2\eminus{10} \pm 2\eminus{11}$} 
	& \cuscmd{$9.3\eminus{10} \pm 1\eminus{11}$} 
	& 
	& 
	\\[4pt]
	& 
	& 1
	& \cuscmd{$9.4\eminus{09} \pm 3\eminus{09}$} 
	& \cuscmd{$8.7\eminus{08} \pm 7\eminus{08}$} 
	& \cuscmd{$8.1\eminus{08} \pm 2\eminus{08}$} 
	& 
	& 
	\\
	& 
	& 5
	& $8.6\eminus{09} \pm 1\eminus{09}$ 
	& \cuscmd{$7.6\eminus{08} \pm 3\eminus{08}$} 
	& \cuscmd{$9.2\eminus{08} \pm 6\eminus{09}$} 
	& 
	& 
	\\
	& 
	& 10
	& $1.1\eminus{08} \pm 4\eminus{09}$ 
	& \cuscmd{$8.3\eminus{08} \pm 3\eminus{08}$} 
	& \cuscmd{$9.0\eminus{08} \pm 4\eminus{09}$} 
	& 
	& 
	\\
	\multirow{-4}{*}{\Prop[3]}
	& \multirow{-4}{*}{\rotatebox{90}{9.00\eminus{08}}}
	& 30
	& $1.3\eminus{08} \pm 5\eminus{09}$ 
	& \cuscmd{$9.1\eminus{08} \pm 2\eminus{08}$} 
	& \cuscmd{$9.1\eminus{08} \pm 3\eminus{09}$} 
	& 
	& 
	\\[4pt]
	& 
	& 1
	& \error 
	& \cuscmd{$6.2\eminus{11} \pm 2\eminus{11}$} 
	& 
	& 
	& 
	\\
	& 
	& 5
	& \error 
	& \cuscmd{$6.0\eminus{11} \pm 6\eminus{12}$} 
	& 
	& 
	& 
	\\
	& 
	& 10
	& \error 
	& \cuscmd{$5.8\eminus{11} \pm 4\eminus{12}$} 
	& 
	& 
	& 
	\\
	\multirow{-4}{*}{\Srop[4]}
	& \multirow{-4}{*}{\rotatebox{90}{5.64\eminus{11}}}
	& 30
	& \error 
	& \cuscmd{$5.5\eminus{11} \pm 2\eminus{12}$} 
	& 
	& 
	& 
	\\[4pt]
	& 
	& 1
	& 
	& \cuscmd{$7.0\eminus{15} \pm 5\eminus{15}$} 
	& \cuscmd{$7.1\eminus{15} \pm 2\eminus{15}$} 
	& 
	&
	\\
	& 
	& 5
	& 
	& $8.3\eminus{15} \pm 4\eminus{15}$ 
	& \cuscmd{$7.7\eminus{15} \pm 1\eminus{15}$} 
	& 
	& 
	\\
	& 
	& 10
	& 
	& $7.2\eminus{15} \pm 2\eminus{15}$ 
	& \cuscmd{$7.7\eminus{15} \pm 6\eminus{16}$} 
	& 
	& 
	\\
	\multirow{-4}{*}{\Srop[5]}
	& \multirow{-4}{*}{\rotatebox{90}{$\approx7.57\eminus{15}$}}
	& 30
	& 
	& $8.8\eminus{15} \pm 3\eminus{15}$ 
	& \cuscmd{$8.3\eminus{15} \pm 4\eminus{16}$} 
	& 
	& 
	\\[4pt]
	& 
	& 1
	& 
	& 
	& \cuscmd{$2.9\eminus{07} \pm 1\eminus{08}$} 
	& $\geqslant 3.0 \eminus{07}$ 
	&
	\\
	& 
	& 5
	& 
	& 
	& \cuscmd{$3.0\eminus{07} \pm 7\eminus{09}$} 
	& \ditto 
	& 
	\\
	& 
	& 10
	& 
	& 
	& \cuscmd{$3.0\eminus{07} \pm 5\eminus{09}$} 
	& \ditto 
	& 
	\\
	\multirow{-4}{*}{\Prop[6]}
	& \multirow{-4}{*}{\rotatebox{90}{$3.06\eminus{07}$}}
	& 30
	& 
	& 
	& \cuscmd{$3.0\eminus{07} \pm 3\eminus{09}$} 
	& \ditto 
	& 
	\\[4pt]
	& 
	& 1
	& 
	& 
	& \cuscmd{$1.7\eminus{07} \pm 4\eminus{08}$} 
	& \cuscmd{$\geqslant 2.8 \eminus{18}$} 
	&
	\\
	& 
	& 5
	& 
	& 
	& \cuscmd{$1.8\eminus{07} \pm 2\eminus{08}$} 
	& \ditto 
	& 
	\\
	& 
	& 10
	& 
	& 
	& \cuscmd{$1.7\eminus{07} \pm 1\eminus{08}$} 
	& \ditto 
	& 
	\\
	\multirow{-4}{*}{\Prop[7]}
	& \multirow{-4}{*}{\rotatebox{90}{$1.70\eminus{07}$}}
	& 30
	& 
	& 
	& \cuscmd{$1.8\eminus{07} \pm 8\eminus{09}$} 
	& \ditto 
	& 
	\\[4pt]
	& 
	& 1
	& 
	& 
	& \cuscmd{$1.5\eminus{06} \pm 3\eminus{07}$} 
	& \error 
	&
	\\
	& 
	& 5
	& 
	& 
	& \cuscmd{$1.5\eminus{06} \pm 1\eminus{07}$} 
	& $\geqslant 2.3 \eminus{28}$ 
	& 
	\\
	& 
	& 10
	& 
	& 
	& \cuscmd{$1.6\eminus{06} \pm 9\eminus{08}$} 
	& \ditto 
	& 
	\\
	\multirow{-4}{*}{\Prop[8]}
	& \multirow{-4}{*}{\rotatebox{90}{$\approx1.20\eminus{06}$}}
	& 30
	& 
	& 
	& \cuscmd{$1.7\eminus{06} \pm 5\eminus{08}$} 
	& \ditto 
	& 
	\\[4pt]
	& 
	& 1
	& $2.2\eminus{04} \pm 6\eminus{06}$ 
	& $2.3\eminus{04} \pm 3\eminus{05}$ 
	& $2.1\eminus{04} \pm 3\eminus{05}$ 
	& 
	& $2.2\eminus{04} \pm 2\eminus{05}$ 
	\\
	& 
	& 5
	& $2.2\eminus{04} \pm 5\eminus{07}$ 
	& $2.2\eminus{04} \pm 1\eminus{05}$ 
	& $2.2\eminus{04} \pm 1\eminus{05}$ 
	& 
	& $2.2\eminus{04} \pm 7\eminus{06}$ 
	\\
	& 
	& 10
	& $2.2\eminus{04} \pm 2\eminus{07}$ 
	& $2.2\eminus{04} \pm 1\eminus{05}$ 
	& $2.2\eminus{04} \pm 1\eminus{05}$ 
	& 
	& $2.2\eminus{04} \pm 5\eminus{06}$ 
	\\
	\multirow{-4}{*}{\Prop[9]}
	& \multirow{-4}{*}{\rotatebox{90}{$2.20\eminus{04}$}}
	& 30
	& $2.2\eminus{04} \pm 1\eminus{07}$ 
	& $2.2\eminus{04} \pm 5\eminus{06}$ 
	& $2.2\eminus{04} \pm 5\eminus{06}$ 
	& 
	& $2.2\eminus{04} \pm 3\eminus{06}$ 
	\\[4pt]
	& 
	& 1
	& $1.1\eminus{05} \pm 8\eminus{06}$ 
	& $8.1\eminus{06} \pm 1\eminus{05}$ 
	& $6.7\eminus{06} \pm 3\eminus{06}$ 
	& 
	& $1.4\eminus{05} \pm 5\eminus{06}$ 
	\\
	& 
	& 5
	& \cuscmd{$1.0\eminus{05} \pm 2\eminus{06}$} 
	& $7.3\eminus{06} \pm 5\eminus{06}$ 
	& $1.2\eminus{05} \pm 2\eminus{06}$ 
	& 
	& $1.0\eminus{05} \pm 2\eminus{06}$ 
	\\
	& 
	& 10
	& $9.7\eminus{06} \pm 2\eminus{06}$ 
	& $1.0\eminus{05} \pm 5\eminus{06}$ 
	& $1.1\eminus{05} \pm 1\eminus{06}$ 
	& 
	& $9.8\eminus{05} \pm 1\eminus{06}$ 
	\\
	\multirow{-4}{*}{\Prop[10]}
	& \multirow{-4}{*}{\rotatebox{90}{$1.00\eminus{05}$}}
	& 30
	& $1.0\eminus{05} \pm 1\eminus{06}$ 
	& $8.9\eminus{06} \pm 2\eminus{06}$ 
	& $9.7\eminus{06} \pm 7\eminus{07}$ 
	& 
	& $1.1\eminus{05} \pm 8\eminus{07}$ 
	\\
	\arrayrulecolor{black}
	\bottomrule
  \end{tabular}
%
%
\end{table}

One or more rare-event properties are given per model:
We used the tools to estimate them, showing the results in \Cref{tab:experiments}.
Per model and property we had the tools run for 1, 5, 10, and 30 minutes (indicated in column \!\raisebox{-1pt}{\larger\clock}) in the TACAS VM~\cite{TacasVm}.
Each tool could use a default run (minimal configuration) or custom commands for that model-property combination.
\Cref{tab:experiments} reports only one of those values: when the difference between them is below 15\% we report the default run; else, we report the one closer to the exact property value~\cite{RareEventsArtifact}.
In the table, an empty cell indicates no support for that property/model.
Values produced by a custom command are marked with a superscript star\,\textsuperscript{$\star$}.
The values reported are either
95\% confidence intervals ({$p\pm\delta$}),
sound lower bounds ($\geqslant p$),
failures ($\varnothing$),
or omitted computations (\!\raisebox{.7ex}{\normalsize\,\rotatebox{90}{-}\hspace{-2pt}\rotatebox{90}{-}}).
The latter applies to \eg model checkers like \ragtimer that use one runtime since longer runs are seldom beneficial.
In general, smaller confidence intervals and results closer to the true value (indicated in the second column under heading ``Prop.'' as either a statistical approximation $\approx p$ or truncated exact value $p$, obtained from reference material or computed with higher resources, viz.\ more memory, runtime, and \acronym{cpu} power) are better.

We note that the default (``sound'') run of \dftres can run longer or shorter than the hard time limit, and its renewal-theory implementation cannot compute \Srop[4] on that \jani model;
also, \fig and \modes used crude Monte Carlo (not \RES) to analyse the DFTs
because no useful importance function could be derived when the dominant failures have short traces;
and \ragtimer used one runtime per property, since longer runs are seldom beneficial for model checkers.

\subsubsection{Data availability.}

We provide an artifact allowing a full experimental reproduction at DOI \href{https://doi.org/10.6084/m9.figshare.23818395}{10.6084/m9.figshare.23818395}~\cite{RareEventsArtifact}.

\subsection{Outlook}
\label{sec:conclu}

We see the need for further research to unify rare event approaches in the formal tools community, \eg to allow \textbf{automatic identification of the algorithm to use}.
A concrete example is the high performance of \dftres (using \IS) to analyse the DFTs in contrast to its comparatively low performance for properties in queueing systems.
This in contrast to \fig and \modes, which (using \ISPLIT) performed well for the latter, but found crude Monte Carlo to be their best approach for the DFTs.

\section{Robust Decision-Making Under Uncertainty}
\label{sec:Uncertainty}

In recent years, there has been a strong push to combine the areas of formal verification---in particular model checking---and artificial intelligence (AI).
A specific area that is native to both of those areas is \textit{decision-making under uncertainty}~\cite{Koc15}.
The level and type of uncertainty affect the capabilities of AI systems to make intelligent decisions.
The core problem is to provide a guarantee that an AI system, operating under uncertainty, adheres to some formally specified constraint, \eg given as a temporal logic specification (see \Cref{sec:LTL}). 
State-of-the-art approaches use models, in particular MDPs, to capture sequential decision-making problems for agents operating in uncertain environments.
Moreover, sensor limitations may lead to partial observability of the system's current state, giving rise to POMDPs (see \Cref{sec:POMDPs}).
MDPs augmented with a model of adversarial behaviour are stochastic games (SGs, see \Cref{sec:StochasticGames}) and their partially observable counterpart is a partially-observable SG (POSG).

The likelihood of uncertain events, such as a message loss in communication channels or specific responses by human operators, may only be an estimate from data.
The models mentioned above capture uncertainty in the form of precise probabilities---either in their transition dynamics or in their observation models.
However, such \emph{point estimates} of probabilities from data carry the risk of statistical errors.
Moreover, the optimal policies for agents are usually highly sensitive to small perturbations in transition probabilities, leading to suboptimal outcomes such as a deterioration in performance~\cite{MSST07,GG23}.

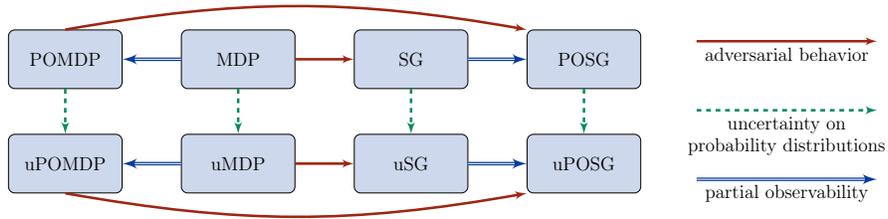
\begin{figure}[t]
\centering
\scalebox{0.6}{%
  	\newcommand{\ndistance}{1.3cm}%
\newcommand{\vdistance}{1.0cm}%
\centering%
\begin{tikzpicture}%
\tikzstyle{all}=[draw, text centered, shape=rectangle, rounded corners, minimum height=1.3cm, fill=mydarkblue!20]%
\tikzstyle{inner}=[text width=2.3cm,font=\large]%
\tikzstyle{split}=[rectangle split,rectangle split parts =2,rectangle split part align=base]%
\tikzstyle{adv}=[ultra thick, draw=myred, font=\large]
\tikzstyle{pobs}=[thick, draw=mydarkblue, double, font=\large]
\tikzstyle{unc}=[ultra thick, draw=mygreen, dashed, font=\large]

\node[all,inner] (mdp) {MDP};%

\node[all,inner,right=\ndistance of mdp] (sg) {SG};%

\node[all,inner,left=\ndistance of mdp] (pomdp) {POMDP};%

\node[all,inner,below=\vdistance of mdp] (umdp) {uMDP};%

\node[all,inner,below=\vdistance of pomdp] (upomdp) {uPOMDP};%

\node[all,inner,right=\ndistance of sg] (posg) {POSG};%

\node[all,inner,below=\vdistance of sg] (usg) {uSG};%

\node[all,inner,below=\vdistance of posg] (uposg) {uPOSG};%


\draw (mdp) edge[-latex', adv] (sg);

\draw (mdp) edge[-latex',pobs] (pomdp);

\draw (mdp) edge[-latex',unc] (umdp);

\draw (sg) edge[-latex', pobs] (posg);
\draw (sg) edge[-latex',unc] (usg);

\draw (pomdp) edge[-latex',unc] (upomdp);

\draw (pomdp.north) edge[-latex',adv, bend left=10] (posg.north west);

\draw (umdp) edge[-latex',pobs] (upomdp);

\draw (umdp) edge[-latex',adv] (usg);

\draw (usg) edge[-latex',pobs] (uposg);

\draw (upomdp.south) edge[-latex',adv, bend right=10] (uposg.south west);

\draw (posg) edge[-latex',unc] (uposg);

\node[right=1cm of posg,yshift=0.4cm] (dummyadv1) {};%
\node[right=4cm of dummyadv1] (dummyadv2) {};%
\draw (dummyadv1) edge[-latex',adv] node[below]{{adversarial behavior}}(dummyadv2);

\node[below=1.3cm of dummyadv1] (dummyunc1) {};%
\node[right=4cm of dummyunc1] (dummyunc2) {};%
\draw (dummyunc1) edge[-latex',unc] node[below,align=center]{uncertainty on \\ probability distributions}(dummyunc2);

\node[below=1.3cm of dummyunc1] (dummypobs1) {};%
\node[right=4cm of dummypobs1] (dummypobs2) {};%
\draw (dummypobs1) edge[-latex',pobs] node[below]{{partial observability}}(dummypobs2);
\end{tikzpicture}%
%
  }
\caption{{A family of closely related uncertainty models}}
  \label{fig:uncertainty-models}
\end{figure}

\emph{Uncertainty models}, sometimes also referred to as \emph{robust models}, remove this assumption by incorporating uncertainty sets of probabilities.
In the literature, uncertain MDPs use, for example, \emph{probability intervals} or \emph{likelihood functions}~\cite{NG05,WKS14}.
Similar extensions exist for uncertain POMDPs, where uncertainty may also affect the observation model~\cite{CJJMST21,SJCT20,BB07,IN07,BR11}.
To the best of our knowledge, there are no results on uncertain POSGs.
\Cref{fig:uncertainty-models} shows a family of \emph{uncertainty models}, capturing different types of uncertainty and their relation to each other.
The three different types of arrows indicate the addition of
(1)~adversarial behaviour,
(2)~uncertainty on probability distributions, and
(3)~partial observability from one model to another.
In the figure, adversarial behaviour increases from left to right.
The left and right columns are partially observable models.
Finally, the bottom row shows models that (in addition to probabilistic and adversarial behaviour) account for uncertainty in probability distributions.
For an overview, we refer the interested reader to \eg \cite{BSSJ23}.

\subsection{Algorithms and Tool Support}

\prism~\cite{KNP11}, a widely-used probabilistic model checker, provides support for two common classes of uncertainty models:
\emph{interval discrete-time Markov chains} (IDTMCs) and \emph{interval Markov decision processes} (IMDPs).
These are specified to the tool with a simple extension of the \prism modelling language where the probabilities attached to variable updates within a guarded command are optionally provided as intervals, for example
$${[}\mathit{move}{]} \ \mathit{loc}=1 \ \rightarrow \ [0.85,0.95]{:}\,(\mathit{loc}'\!=2) + [0.05,0.15]{:}\,(\mathit{loc}'\!=1){;}\footnote{In this example, $\mathit{move}$ labels the transitions induced by the command, $\mathit{loc}=1$ is the guard that determines when the command is enabled, and each of the two branches to the right of $\rightarrow$ has an interval of probabilities and a set of assignments.}$$
and, as usual, can be given as expressions in terms of variables and parameters:
$${[}\mathit{send}{]} \ \mathit{s}\!=\!1 \ \!\rightarrow\! \ [p_\mathit{fail}-\varepsilon,p_\mathit{fail}+\varepsilon]{:}\,(\mathit{s}'\!\!=\!1) + [(1{-}p_\mathit{fail})-\varepsilon,(1{-}p_\mathit{fail})+\varepsilon]{:}\,(\mathit{s}'\!\!=\!2);$$
This makes it straightforward to adapt existing DTMC or MDP benchmarks~\cite{KNP12} to their interval variants, as done for example in~\cite{PLSS13}.

\prism provides \emph{robust} verification, quantifying over all possible transition probabilities contained within the models' uncertainty sets.
Property specification extends the existing \prism property language.
For IDTMCs and IMDPs, the tool supports the temporal logic PCTL, extended with (expected) reward operators and (co-safe) LTL formulae.
For example, formulas
${\tt P}_{\max\!\min=?}[\,{\tt F}\,\mathit{goal}\,]$ and ${\tt P}_{\max\!\max=?}[\,{\tt F}\,\mathit{goal}\,]$
ask for the worst- and best-case scenarios, respectively, for maximising the probability of reaching a $\mathit{goal}$-labelled state.

Like many probabilistic model checking implementations, the uncertain models are solved via dynamic programming, in this case, \emph{robust value iteration}~\cite{NG05,WTM12},
implemented in \prism's Java-based explicit-state model checking engine.
Optimal policies for IMDPs can be generated and exported or simulated.
Access to IDTMC and IMDP model checking is also provided programmatically at an API level, and has been applied to various problems, including anytime model learning~\cite{SSPJ22} and abstraction of dynamical systems~\cite{BRAPPSJ23}.

\subsection{Outlook}

Tool support for uncertainty models can be extended in various directions,
for example to provide model checking for some of the model classes identified in \Cref{fig:uncertainty-models} featuring partial observability (uncertain POMDPs) or adversarial behaviour (uncertain SGs), as well as improving efficiency and scalability for the simpler model classes.
It will also be beneficial to extend the range of uncertainty types beyond intervals, which also necessitates more significant modelling language extensions.

\section{State Space Exploration}
\label{sec:StateSpaceExploration}

State space exploration engines form the foundation of numerous quantitative analysis tools, playing a pivotal role in their functionality.
Explicit-state model checkers, such as \storm with its \texttt{sparse} engine and \mcsta, rely on exploration engines to exhaustively construct the complete state space of a model before applying probabilistic model checking algorithms.
Additionally, statistical model checkers such as \modes leverage exploration engines to generate large amounts of traces for statistical analysis.
Exploration engines have recently also been used for training and verifying reinforcement learning agents~\cite{GJJP22,GHHKKW22}.

In an effort to better understand the performance characteristics of the exploration engines utilised in different tools, we systematically benchmark and compare them.
For this purpose, we consider the time and space needed for building an explicit representation of the complete state space of a model.
Additionally, we compare the engines based on qualitative criteria such as the types of models they can handle and the interfaces they provide.

\subsection{Tool Support}

The tools participating in this category are the \toolset, \momba, and \storm.
Both \momba and \storm participate with multiple engines, adding further diversity to the evaluation.
Since all three tools support \jani, we employ it as a foundation for comparing and contrasting their capabilities.
\begin{description}[itemsep=3pt,topsep=3pt,labelsep=0pt]

\item[The \toolset]
includes a state space exploration engine written in C\# that is used by several of its tools, including \mcsta and \modes.
It supports all types of models specified by \jani, including all \jani extensions.
In that regard, it stands out as the most versatile among the engines we consider.
For PTA, the engine supports the digital clock semantics~\cite{KNPS06}, explicit valuations, clock regions~\cite{HSD17}, as well as clock zones~\cite{DHLS16}.
In addition, it supports a symbolic treatment of continuous variables for hybrid models.
In contrast to both \storm and \momba, which both provide public interfaces to their engines, the \toolset's engine is intended for internal use only and does not provide a public interface.
The \toolset includes a separate \tool{mopy} transpilation tool to convert models to Python code implementing a first-state-next-state interface which can be used to explore the model's state~space.
In our experiments below, we access the \toolset's state space exploration engine via \mcsta.

\item[\momba]
includes as a key feature a state space exploration engine designed to make exploration readily accessible via a comprehensive Python API.
To achieve good performance, the engine is written in Rust.
While \momba itself supports all of \jani, its state space exploration engine is more limited:
It supports all discrete-time model types and flavours of timed automata specified by \jani except stochastic timed automata.
The supported \jani extensions are \texttt{arrays}, \texttt{derived-operators}, \texttt{named-expressions}, and \texttt{trigonometric-functions}.
In particular, the \texttt{functions} extension is not supported yet.
For timed automata, it supports explicit valuations as well as clock zones.
The Python API also provides functionality that goes beyond mere exploration:
for instance, arbitrary \jani expressions can be evaluated in a given state and, for timed automata, clock zones can be manipulated.
In addition to its traditional state space exploration engine, \momba also participates with an experimental new engine supporting a parallelized exploration mode harnessing the potential of multi-core systems.
This experimental engine does not currently support timed automata and is not yet exposed via the Python API.

\item[\storm]
participates with its \texttt{sparse} and \texttt{dd-to-sparse} engines.
While \storm's \texttt{sparse} engine, like the engines of the \toolset and \momba, adopts a conventional explicit-state approach, the \texttt{dd-to-sparse} engine is based on first constructing a symbolic representation using BDDs of the state space and subsequently translating this to a traditional explicit representation.
\storm supports all discrete- and continuous-time model types specified by \jani, except timed and hybrid automata.
The supported \jani extensions are \texttt{arrays}, \texttt{derived-operators}, \texttt{functions}, and \texttt{state-exit-rewards}.
\storm provides both a C\texttt{++} and a Python interface, the latter as part of \tool{Stormpy}, to its state space exploration engine.
While the C\texttt{++} API is fully featured, the Python API only supports the exploration of the entire state space of \jani models (but not the simulation of individual traces) while it has no such limitation for \prism models.
In contrast to Modest Toolset and \momba, \storm offers support for arbitrary-precision arithmetic using rational numbers implemented in the GMP library. 
This enables precise calculations and analysis, particularly when dealing with models that require high precision.

\end{description}

\subsection{Performance Comparison}

In our experimental evaluation, we utilise the QVBS as the foundation for benchmarking the tools.
To ensure a meaningful comparison, we focus exclusively on discrete-time models, as these are supported by all the participating tools.
Out of our initial 229 QVBS benchmarks, 25 resulted in timeouts after 30 minutes or were unsupported by all tools.
Hence, the following analysis focuses on the remaining 204 benchmarks.
For each benchmark, we measure the time required by each state space exploration engine to construct the entire state space.
Additionally, we track the number of states counted by the engines and assess the memory consumption associated with each state where applicable.
All benchmarks ran on a computer equipped with a 16-core AMD EPYC-Milan processor running at 3.4\,GHz and 128\,GB of RAM.

\begin{table}[t]
\caption{Number of benchmarks per outcome and state space exploration engine}
\label{tb:outcomes}
\centering
\setlength{\tabcolsep}{6pt}
\renewcommand*{\arraystretch}{1.1}
\begin{tabular}{lcccc}
    \toprule
    Engine & \textit{solved} & \textit{unsupported} & \textit{timeout} & \textit{error} \\\midrule
    \toolset & 194 & 9 & 1 & 0\\
    \momba (v1) & 159 & 45 & 0 & 0 \\
    \momba (v2,\texttt{seq}) & 159 & 45 & 0 & 0\\
    \momba (v2,\texttt{par}) & 154 & 45 & 5 & 0\\
    \storm (\texttt{dd-to-sparse}) & 195 & 3 & 2 & 4\\
    \storm (\texttt{sparse})& 202 & 0 & 2 & 0\\\bottomrule
\end{tabular}
\end{table}

Table~\ref{tb:outcomes} shows the number of benchmarks per tool and our experiments' qualitative outcomes:
we display the number of benchmarks that were successfully \textit{solved}, \textit{unsupported}, lead to a \textit{timeout}, or caused an \textit{error}.
The $9$ benchmarks not supported by the \toolset's engine use a complex specification for the initial states.
The $45$ benchmarks not supported by \momba use the \texttt{functions} \jani extension and are a superset of the $9$ benchmarks not supported by the \toolset.
The $3$ benchmarks not supported by \storm's $\texttt{dd-to-sparse}$ engine use assignment indices while for $4$ benchmarks the same engine returned an error due to the BDD implementation running out of memory.
The timeouts are all for different benchmarks.
While the number of states reported by \storm and \momba is the same for all benchmarks and engines, the \toolset sometimes reports fewer states which presumably is due to some state space-reducing optimizations.

\begin{figure}[tp]
\centering
\includegraphics{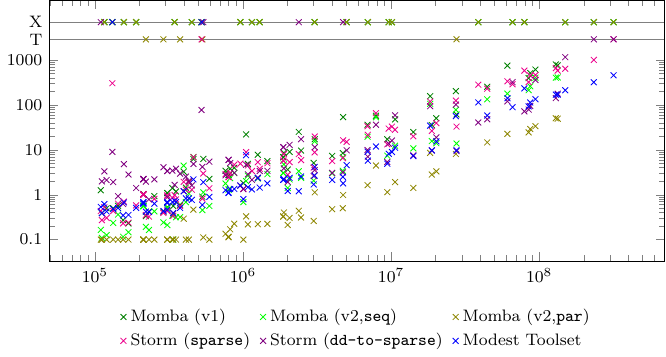}
\caption{Runtimes in seconds in relation to the total number of states}
\label{fig:scatter-plot}
\end{figure}

\begin{figure}[tp]
\centering
\includegraphics{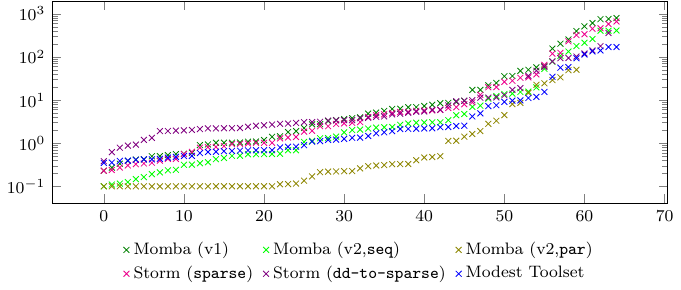}
\caption{Runtimes (s) vs.\ the number of benchmarks each solved in that time}
\label{fig:quantile-plot}
\end{figure}

\paragraph{Runtimes.}
\Cref{fig:scatter-plot} depicts the running time for each benchmark (on the vertical axis) in relation to the total number of states of the respective benchmark (on the horizontal axis).
The marks at the top indicate timeouts (T), and unsupported benchmarks as well as benchmarks returning an error (X).
The quantile plot in \Cref{fig:quantile-plot} presents the cumulative number of benchmarks solved within a certain time.
For presentation purposes, we chose to clamp the running times at $0.1\,s$ and restrict the plots to benchmarks with more than $10^5$ states.
For smaller benchmarks, the differences in runtimes are practically insignificant.
Additionally, \Cref{fig:quantile-plot} is restricted to benchmarks supported by all engines to prevent skewing the plot (as otherwise an unsupported benchmark and a timeout would have the same effect).

From these results, it is evident that the approach taken by the \texttt{dd-to-sparse} engine of \storm only pays off for larger models; even then, it is rarely faster than the conventional explicit engine of the \toolset.
Among those engines exclusively using a single core, the \toolset engine is almost always the fastest, although it has a larger startup overhead.
This does not come as a surprise because, for efficiency, it is based on compiling \jani models to C\# bytecode that is JIT-compiled.
\storm's \texttt{dd-to-sparse} engine, like \momba's experimental parallel engine (v2,\texttt{par}), uses multiple cores since the underlying BDD implementation in \tool{Sylvan}~\cite{DP17} is parallelised.
\momba's parallel engine is always faster than any other engine for benchmarks of a significant size.
The average speed-up when compared to its sequential version is a factor of $9.1$.
In general, though, the runtimes of all engines are often quite similar.

Note that, as \storm and \mcsta are model checkers, they do a bit more work than \momba by creating a sparse matrix representation of the transitions and computing atomic propositions.
We expect the performance impact of this to be minor---however we did not measure it.

\begin{figure}[tp]
\centering
\includegraphics{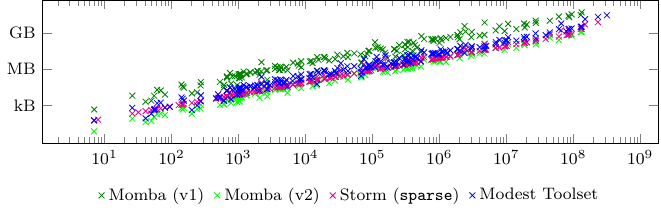}
\caption{Size of the state space in relation to the number of states}
\label{fig:sizes}
\end{figure}

\paragraph{Memory consumption.}
Another interesting dimension when it comes to state space construction is the required memory.
Efficiency is crucial here given the often huge state spaces due to the state space explosion problem.
For the traditional explicit state engines, the size of the state space is linear in the number of states.
\Cref{fig:sizes} shows the size of the state spaces in relation to the number of states, computed based on the number of states and the size of each state.
Note that the sequential and parallel variant of \momba's experimental engine use the same representation.
In contrast to the \toolset, \storm's \texttt{sparse} engine and \momba's experimental engine use a more space efficient bit-packing representation of states, thereby reducing the amount of required memory.
\momba's original engine uses the worst representation and always requires at least $16$ bytes per variable independent of its actual domain.

\paragraph{Summary.}
Our results show that all engines are roughly comparable with respect to the time it takes to construct the entire state space of a model.
\storm's \texttt{dd-to-sparse} engine may only be advantageous in terms of runtime for some large models while incurring a high overhead for small models.
Among single-core engines, the \toolset's engine is almost always the fastest, especially for large models, while being the most versatile at the same time.
The experimental parallel engine of \momba demonstrates that parallel state space exploration can be highly beneficial for larger models.
The original \momba engine requires significantly more memory than all others.
The \toolset's engine, however, does not provide a public API.
Thus, if integration into another tool is a concern, \storm and, in particular, \momba with its original engine have an advantage as they both provide a Python API in addition to APIs in C\texttt{++} and Rust, respectively.

\paragraph{Limitations.}
One of the motivations of this category is the lack assessment for simulation of individual traces.
Note that the performance characteristics displayed here may not carry over to simulation of individual traces as there is a difference between always computing all successor states, as required for exhaustive exploration, and selectively computing only individual successor states which is, for instance, explicitly supported by \momba.
Additionally, an exhaustive exploration requires maintaining a (hash) set of all visited states.

\subsubsection{Data availability.}
An artifact allowing to reproduce the performance comparison is archived and available at DOI \href{https://doi.org/10.5281/zenodo.10626177}{10.5281/zenodo.10626177}~\cite{StateSpaceExplorationArtifact}.

\section{Stochastic Games}
\label{sec:StochasticGames}

Game theory provides an effective way to model strategic interactions between multiple agents (or players) collaborating or competing to achieve objectives.
Games have long been of interest within formal verification, providing a natural way to model, for example, honest and malicious participants in a security protocol or a controller operating in an adversarial environment.

In the context of quantitative verification, \emph{stochastic games} (SGs) are a natural model to reason about strategic interactions in the context of uncertainty, 
\begin{wrapfigure}[16]{r}{0.36\linewidth}
\centering\vspace{-2.2em}
\includegraphics[width=0.32\textwidth, clip, trim=0.0cm 5.9cm 0.0cm 0.0cm]{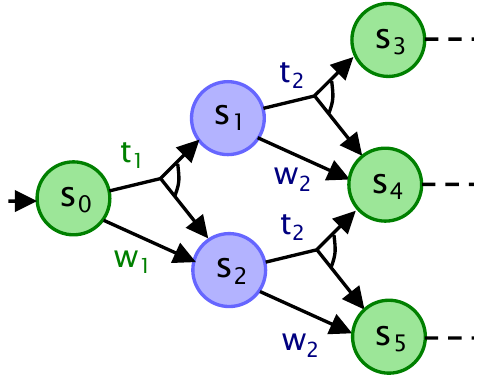}\\
{\scriptsize (a) TSG}\\[6pt]
\includegraphics[width=0.32\textwidth, clip, trim=0.0cm 0.0cm 0.0cm 7.8cm]{figures/stochastic-games/tsg-csg}\\
{\scriptsize (b) CSG}\\
\end{wrapfigure}
noise, or randomisation.
Verification problems for SGs have been studied for over 20 years~\cite{CH12};
the first model checking tools for SGs appeared over 10 years ago~\cite{CFKPS12}, and there has been growing interest in the topic recently.

In essence, SGs (visualised on the right) generalise MDPs by permitting multiple players to have distinct strategies about how to resolve nondeterministic choices in the model.
The simplest model, a turn-based SG (TSG), simply partitions the state space of an MDP, with the choices in each state being under the control of exactly one player.
A concurrent SG (CSG) provides a more realistic model of concurrent decision-making:
in each state, players resolve their choices independently.

Verification of SGs also takes a variety of flavours.
The simplest option is a \emph{zero-sum} setting, where one player (or a coalition of players) aims to maximise some objective, such as the probability of reaching a set of target states or satisfying a temporal logic formula, and the other player (or players) have the opposite objective, \ie to minimise it.
For SGs, the logic rPATL~\cite{CFKPS12} is widely used, which generalises the well-known game logic ATL~\cite{AHK02} to a variety of quantitative objectives.
Beyond zero-sum properties, temporal logics and model checking algorithms have been extended~\cite{KNPS21} to support \emph{equilibria}, which are joint strategies where each player optimises their own distinct objective in such a way that it is not beneficial for any player to unilaterally change strategy.

\subsection{Algorithms and Tool Support}

Despite verification problems for SGs typically having a higher complexity than their MDP counterparts, core properties of TSGs can be effectively analysed with similar methods such as value iteration~\cite{Con90} or interval iteration and its variants~\cite{EKKW22,AEKSW22}.
Methods to solve CSGs tend to be more expensive:
again they are usually based on value iteration, but require the solution of a linear programming or equilibrium synthesis problem~\cite{KNPS21} for every state at each iteration.

Verification tools for SGs under active development are
\prismgames and its extensions, \tempest, \pet, and \epmc.
We provide a brief empirical comparison of the first four below.
These tools share a common input format for SGs, namely the \prismgames modelling language.
This extends the widely used \prism modelling language:
In the case of TSGs, it is a rather simple extension of the case for MDPs, defining a set of players and the states they own;
CSGs use a different model of parallel composition and additional language features.
\begin{description}[itemsep=3pt,topsep=3pt,labelsep=0pt]

\item[\epmc]
also supports the analysis of stochastic parity games and verification of epistemic properties on probabilistic multi-agent systems in addition to its standard probabilistic model checking functionality.

\item[\pet]
has recently been extended to support reachability objectives for TSGs.
It uses \prismgames to parse and explore games, and employs the interval iteration approach of~\cite{EKKW22} to solve them.
Implementing partial exploration based on~\cite[Sect.~5]{EKKW22} in combination with the approach of~\cite{KMW23} for more complex objectives such as total reward or mean payoff is planned.

\item[\prismgames]
mainly focuses on TSGs and CSGs, but it also supports turn-based probabilistic timed games.
The tool supports a wide range of zero-sum properties (probabilistic reachability, expected rewards, co-safe linear temporal logic and multi-objective specifications) as well as (social welfare) Nash equilibria.
Recent extensions add support for correlated equilibria and social fairness~\cite{KNPS22a}.
The implementation is primarily based on variants of value iteration, implemented in Java with explicit state data structures, but also includes symbolic (MTBDD-based) model checking of TSGs~\cite{KNPS22b}.

\item[\tempest]
extends \storm to TSGs with a focus on synthesizing most-permissive strategies.
The tool supports zero-sum properties, namely probabilistic reachability and mean-payoff properties.
The model checking procedures are based on variants of value iteration using explicit representations of the state space.

\item[\cite{KRSW22} and \cite{AEKSW22}~]
present an extension of \prismgames which adds various methods for solving TSGs:
interval iteration (II)~\cite{EKKW22} and optimistic value iteration (OVI)~\cite{AEKSW22}, as well as topological variants of each;
the ``widest path'' (WP) variant of II~\cite{PTHH20};
and solution methods based on strategy iteration and quadratic programming.
The latter are omitted from our comparison since they are fundamentally different from the variants of value iteration employed by the other tools \cite[Sect.~5.5.3]{KRSW22}; we refer to~\cite[Sect.~5]{KRSW22} for a practical comparison of these solution methods.

\end{description}
Also relevant are \gist~\cite{CHJR10} and \gavs~\cite{CKLB11}, which implement TSG verification, but are no longer developed or maintained, and \tool{Uppaal Stratego}~\cite{DJLMT15}, which supports stochastic priced timed games via multiple other \tool{Uppaal} branches.

\subsection{Performance Comparison}

\begin{table}[t]
\centering
\caption{Performance comparison results of tools for stochastic games}
\label{tab:perf-sg}
\scriptsize
\setlength{\tabcolsep}{2pt}
\renewcommand*{\arraystretch}{1.2}
\begin{tabular}{ccrrrrrrrr} \toprule
\multicolumn{3}{c}{Benchmark} &
\multicolumn{3}{c}{Value iteration (s)} &
\multicolumn{4}{c}{$\varepsilon$-exact (s)} \\
\cmidrule{1-3}\cmidrule(l){4-6}\cmidrule(l){7-10}
\multirow{3}{*}{\shortstack[c]{Model + property \\ {[}parameters{]}}} &
\multirow{3}{*}{\shortstack[c]{Param. \\ values}} &
\multicolumn{1}{c}{\multirow{3}{*}{\shortstack[c]{\#\,states}}} &
\multicolumn{1}{c}{\multirow{3}{*}{\shortstack[c]{~\tool{Prism}\\ ~\tool{-games}\\~(expl.)}}} &
\multicolumn{1}{c}{\multirow{3}{*}{\shortstack[c]{\tool{Prism} \\ \tool{-games} \\ (symb.)}}} &
\multicolumn{1}{c}{\multirow{3}{*}{\shortstack[c]{\tool{Temp} \\ -\tool{est}}}} &
\multicolumn{1}{c}{\multirow{3}{*}{\shortstack[c]{~\pet}}} &
\multicolumn{1}{c}{\multirow{3}{*}{\shortstack[c]{\tool{P-g+} \\ (II)}}} &
\multicolumn{1}{c}{\multirow{3}{*}{\shortstack[c]{\tool{P-g+} \\ (OVI)}}} &
\multicolumn{1}{c}{\multirow{3}{*}{\shortstack[c]{\tool{P-g+} \\ (WP)}}} \\
&&&&&&&&&\\
&&&&&&&&&\\
\midrule
\multirow{3}{*}{\shortstack[c]{\emph{avoid}\,+\,\emph{find} \\ $[\mathtt{X{\uu}MAX},\mathtt{Y{\uu}MAX}]$}}
& 10,\,10 & 106,524 & 16.9 & 15.4 & \sgbest{1.4} & \sgbest{5.0} & 17.2 & 22.4 & 16.7 \\
& 15,\,15 & 480,464 & 125.9 & 62.6 & \sgbest{4.7} & \sgbest{15.7} & 126.9 & 137.2 & 126 \\
& 20,\,20 & 1,436,404 & \sgtimeout & 240.8 & \sgbest{12.9} & \sgbest{57.5} & \sgtimeout & \sgtimeout & \sgtimeout \\[3pt]
\multirow{3}{*}{\shortstack[c]{\ \emph{hallway{\uu}human}\\+\,\emph{save} \\ $[\mathtt{X{\uu}MAX},\mathtt{Y{\uu}MAX}]$}}
& 5,\,5 & 25,000 & 2.5 & 1.8 & \sgbest{0.9} & 2.9 & \sgbest{2.4} & \sgbest{2.4} & \sgbest{2.4} \\
& 10,\,10 & 400,000 & 10.5 & \sgbest{2.0} & 12.9 & \sgbest{9.5} & 11.3 & 11.2 & 11.3 \\
& 15,\,15 & 2,025,000 & 50.1 & \sgbest{4.0} & 101.3 & \sgbest{39.6} & 57.0 & 55.4 & 56.6 \\[3pt]
\multirow{4}{*}{\shortstack[c]{\emph{investors}\,+\,\emph{greater} \\ $[\mathtt{N},\mathtt{vmax}]$}}
& 2,\,20 & 568,790 & 21.8 & 7.4 & \sgbest{4.9} & \sgbest{16.7} & 33.2 & 42.3 & 54.6 \\
& 2,\,40 & 2,041,690 & 98.8 & 26.0 & \sgbest{19.8} & \sgbest{69.0} & 144.8 & 183.2 & 314.6 \\
& 3,\,20 & 4,058,751 & 167.7 & \sgbest{19.2} & 39.7 & ~\sgbest{152.6} & 241.4 & 321.3 & 484.8 \\
& 3,\,40 & 14,569,251 & \sgmemout & \sgbest{62.8} & 171.2 & \sgtimeout & \sgtimeout & \sgtimeout & \sgtimeout \\[3pt]
\multirow{2}{*}{\shortstack[c]{\emph{safe{\uu}nav}\,+\,\emph{reach} \\ $[\mathtt{N},\mathtt{feat}]$}}
& 8,\,D & 2,592,845 & ~544.2 & \sgbest{16.2} & 518.5 & 519.4 & 498.4 & 508.7 & \sgbest{485.7} \\
& 8,\,A & 17,052,941 & \sgtimeout & \sgbest{110.8} & \sgtimeout & \sgtimeout & \sgtimeout & \sgtimeout & \sgtimeout \\[3pt]
\multirow{2}{*}{\shortstack[c]{\emph{BigMec}\,+\,\emph{BigMec}\\ $[\mathtt{N}]$}}
& 10,000 & 20,003 & 46.9 & 9.3 & \sgbest{2.5} & \sgbest{17.6} & \sgtimeout & 49.5 & 73.5 \\
& 25,000 & 50,003 & 290.4 & 45.8 & \sgbest{12.7} & \sgbest{82.9} & \sgtimeout & 294.2 & 472.4 \\[3pt]
\multirow{2}{*}{\shortstack[c]{\emph{ManyMec}\,+\,\emph{ManyMec} \\ $[\mathtt{N}]$}}
& 10,000 & 30,002 & 160.7 & 263.3 & \sgbest{16.7} & \sgbest{104.3} & \sgtimeout & \sgtimeout & 460.4 \\
& 25,000 & 75,002 & \sgtimeout & \sgtimeout & \sgbest{98.6} & \sgtimeout & \sgtimeout & \sgtimeout & \sgtimeout \\[1pt]
\bottomrule
\end{tabular}
\end{table}

We give a brief performance comparison of the various tools and techniques, focusing on the problem class supported by all tools:
zero-sum probabilistic reachability for TSGs.
\Cref{tab:perf-sg} shows total runtimes (game construction and solution) on an indicative set of benchmarks from the PRISM Benchmark Suite~\cite{KNP12} and~\cite{KRSW22}.
Experiments ran on an AMD Ryzen 5 3600 system, pinned to a single core and restricted to 8\,GB of RAM, running inside Docker, using OpenJDK JRE-17 for all Java tools, and with a timeout (T/O) of 10 minutes.
For each invocation, a fresh docker container is created.

For a fair comparison, we group them into two distinct categories based on the degree of accuracy provided:
``value iteration'' (\ie no strict guarantees on the correctness of the result) and ``$\varepsilon$-exact'' (the result is guaranteed to be within $\pm\,\varepsilon = 10^{-6}$ of the true value), marking the fastest tool in each category in bold.

\paragraph{Value iteration.}
Comparing explicit-state implementations, \tempest is faster than \prismgames on almost all instances (primarily, it appears, due to the former being implemented in C++, but the latter also uses slower but more extensive precomputations).
\prismgames, in symbolic mode, outperforms \tempest on most larger models and scales to the biggest TSGs of all tools.
Symbolic model building times (not shown) are also usually faster.

\paragraph{$\varepsilon$-exact.}
\pet outperforms the approaches in the \prismgames extension of \cite{KRSW22,AEKSW22} (denoted \tool{P-g+} in the table) on practically all models.
This is interesting since the algorithmic approach in the former is the same as interval iteration (II) in the latter.
Since these tools are implemented in the same language (Java) and use the same model construction (\prism's model generator), the (significant) differences are solely a result of engineering.
Times for the methods in the \prismgames extension are typically in the same order of magnitude, however there are models where one approach significantly outperforms all others.

\subsubsection{Data availability.}
All tools, models and scripts needed to replicate our results can be found at DOI \href{https://doi.org/10.5281/zenodo.7831387}{10.5281/zenodo.7831387}~\cite{StochasticGamesArtifact}.

\subsection{Outlook}

Interest has grown in the formal verification of SGs in recent years and it has already been applied to a range of domains, from computer security to adaptive software architectures (as evidenced by the collection of \prismgames case studies at \href{https://www.prismmodelchecker.org/games/casestudies.php}{prismmodelchecker.org/games/casestudies.php}).
In addition to improving the efficiency and scalability of existing tools, one key challenge is to develop methods for \textbf{partially observable} variants of SG models.
Another is to develop support for \textbf{richer specification languages}, for example incorporating strategies, equilibria or epistemic properties.

\section{Conclusion}
\label{sec:Conclusion}

We have described the state of the art in tools and algorithms at the frontiers of quantitative verification in ten different categories, covering 19 different tools.
In several categories, we reported on the first systematic performance comparison among the included tools.
On many of the frontiers we described, tool support for advanced properties and models is now being consolidated, but a plethora of open questions and unimplemented ideas remain for future work.
We hope that this report can serve as an inspiration for further work on quantitative verification tooling, and that several of QComp 2023's categories can evolve into regular, serious performance evaluations among competing tools in the near future.
At the same time, it is clear that our coverage of the quantitative verification frontiers is not complete.
As one example, we mention the area of parametric models based on timed automata (in which parameters are traditionally more structural in nature than the ones in the parametric Markov models of \Cref{sec:Parameters}) where tools are maturing~\cite{And21} and benchmark sets with support for \jani are being collected~\cite{AMP21}, laying the foundations for future performance evaluations.

For the next edition of QComp, which at the latest will take place in time for the next edition of the TOOLympics, we intend to keep the multiple-category setup.
We plan to both add new categories, \eg on parametric timed automata as mentioned above or on entirely new problems that surface in the coming years, and also perform more extensive performance evaluations in those categories where tools will have matured sufficiently and a good benchmark set will have become available.
As such, we expect a mix of ``friendly'' categories that stimulate tool development and standardisation as well as more ``competitive'' evaluations where performance really counts.
Practically, we may need to split off the reports of the larger categories---those where many tools are evaluated on comprehensive benchmark sets to obtain representative performance comparisons---from the main competition report into publications of their own.
In parallel to the transformation of QComp that started with this edition, the comparison of established tools on basic problems as in QComp 2019 and 2020 is likely to transition into a continuous evaluation---rather than periodic competitions---hosted on the \href{https://qcomp.org/}{qcomp.org} website.
We look forward to a continuing journey into the undiscovered country beyond today's frontiers of quantitative verification in the next editions of the QComp friendly competition!

\begin{table}[t]
\caption{Data availability for QComp 2023}
\label{tab:DataAvailability}
\centering
\setlength{\tabcolsep}{6pt}
\renewcommand*{\arraystretch}{1.2}
\begin{tabular}{cllc} \toprule
Section                         & Category                  &  DOI & Ref.\\\midrule
\ref{sec:LRA}                   & Long-Run Average Rewards  & \href{https://doi.org/10.5281/zenodo.8219191}{10.5281/zenodo.8219191} & \cite{LRAArtifact} \\
\ref{sec:MultiObjective}        & Multi-Objective Analysis  & \href{https://doi.org/10.5281/zenodo.8063883}{10.5281/zenodo.8063883} & \cite{MultiObjectiveArtifact} \\
\ref{sec:Parameters}            & Parametric Markov Models  & \href{https://doi.org/10.5281/zenodo.10646479}{10.5281/zenodo.10646479} & \cite{ParametersArtifact} \\
\ref{sec:POMDPs}                & Partially-Observable MDPs & \href{https://doi.org/10.5281/zenodo.8215337}{10.5281/zenodo.8215337} & \cite{PomdpsArtifact} \\
\ref{sec:RareEvents}            & Rare Events               & \href{https://doi.org/10.6084/m9.figshare.23818395}{10.6084/m9.figshare.23818395} & \cite{RareEventsArtifact} \\
\ref{sec:StateSpaceExploration} & State Space Exploration   & \href{https://doi.org/10.5281/zenodo.10626177}{10.5281/zenodo.10626177} & \cite{StateSpaceExplorationArtifact} \\
\ref{sec:StochasticGames}       & Stochastic Games          & \href{https://doi.org/10.5281/zenodo.7831387}{10.5281/zenodo.7831387} & \cite{StochasticGamesArtifact} \\
\bottomrule
\end{tabular}
\end{table}

\subsubsection{Data availability.}
In each category that performed a performance comparison, we provide an artifact that archives the models, tools, scripts, and other data that is necessary to reproduce the respective experiments.
The benchmark set of parametric Markov models introduced in \Cref{sec:Parameters} is also publicly archived.
We link to the DOIs of the respective datasets at the end of each of sections \ref{sec:LRA}, \ref{sec:MultiObjective}, \ref{sec:Parameters}, \ref{sec:POMDPs}, \ref{sec:RareEvents}, \ref{sec:StateSpaceExploration}, and \ref{sec:StochasticGames}, and list all of them in \Cref{tab:DataAvailability}.

\bibliography{paper}
\bibliographystyle{splncs04}

\end{document}